\documentclass[12pt]{article}

\usepackage{amsmath}
\allowdisplaybreaks
\usepackage{amssymb}

\begin{document}

\baselineskip=16.7pt plus 0.2pt minus 0.1pt

\makeatletter
\@addtoreset{equation}{section}
\renewcommand{\theequation}{\thesection.\arabic{equation}}
\newcommand{\bm}[1]{\boldsymbol{#1}}
\newcommand{\calA}{\mathcal{A}}
\newcommand{\calB}{\mathcal{B}}
\newcommand{\calC}{\mathcal{C}}
\newcommand{\calE}{\mathcal{E}}
\newcommand{\calP}{\mathcal{P}}
\newcommand{\calM}{\mathcal{M}}
\newcommand{\calN}{\mathcal{N}}
\newcommand{\calV}{\mathcal{V}}
\newcommand{\calK}{\mathcal{K}}
\newcommand{\calF}{\mathcal{F}}
\newcommand{\calG}{\mathcal{G}}
\newcommand{\calH}{\mathcal{H}}
\newcommand{\calT}{\mathcal{T}}
\newcommand{\calU}{\mathcal{U}}
\newcommand{\calY}{\mathcal{Y}}
\newcommand{\calW}{\mathcal{W}}
\newcommand{\calL}{\mathcal{L}}
\newcommand{\calD}{\mathcal{D}}
\newcommand{\calO}{\mathcal{O}}
\newcommand{\calI}{\mathcal{I}}
\newcommand{\calQ}{\mathcal{Q}}
\newcommand{\calS}{\mathcal{S}}
\newcommand{\QB}{Q_\textrm{B}}
\newcommand{\nn}{\nonumber}
\newcommand{\drv}[2]{\frac{d #1}{d#2}}
\newcommand{\veps}{\varepsilon}
\newcommand{\eps}{\epsilon}
\newcommand{\ds}{\displaystyle}
\newcommand{\Ke}{K_{\veps}}
\newcommand{\invKe}{\frac{1}{\Ke}}
\newcommand{\Ue}{U_{\veps }}
\newcommand{\Ge}{G_{\veps }}
\newcommand{\Psie}{\Psi_{\veps}}
\newcommand{\CR}[2]{\left[#1,#2\right]}
\newcommand{\ACR}[2]{\left\{#1,#2\right\}}
\newcommand{\Pmatrix}[1]{\begin{pmatrix} #1 \end{pmatrix}}
\newcommand{\tr}{\mathop{\rm tr}}
\newcommand{\Tr}{\mathop{\rm Tr}}
\newcommand{\p}{\partial}
\newcommand{\wh}[1]{\widehat{#1}}
\newcommand{\wt}[1]{\widetilde{#1}}
\newcommand{\ol}[1]{\overline{#1}}
\newcommand{\abs}[1]{\left| #1\right|}
\newcommand{\VEV}[1]{\left\langle #1\right\rangle}
\newcommand{\Drv}[2]{\frac{\p #1}{\p #2}}
\newcommand{\ket}[1]{| #1 \rangle}
\newcommand{\bra}[1]{\langle #1 |}
\newcommand{\braket}[2]{\langle #1 | #2 \rangle}
\newcommand{\KBc}{K\!Bc}
\newcommand{\Ngh}{N_\textrm{gh}}
\newcommand{\Gm}{\Gamma}
\newcommand{\pF}{\calF}
\newcommand{\id}{\mathbb{I}}
\renewcommand{\Re}{\mathop{\rm Re}}
\renewcommand{\Im}{\mathop{\rm Im}}
\newcommand{\Res}{\mathop{\rm Res}}
\newcommand{\Ds}{\Delta_s}
\newcommand{\da}{\frac{2\pi i}{s}}
\newcommand{\Da}{\frac{2\pi i}{s}}
\newcommand{\opz}[1]{\left(1+z #1\right)}
\newcommand{\wpopz}[1]{\left(w+1+z #1\right)}
\newcommand{\whwpz}[1]{\left(\wh{z}+ #1\right)}
\newcommand{\CRBc}{{\mathcal X}}
\newcommand{\dg}{\ddagger}

\makeatother
\begin{titlepage}

\title{
\hfill\parbox{3cm}{\normalsize KUNS-2746}\\[1cm]
{\Large\bf
  Analytic Construction of Multi-brane Solutions in Cubic String
  Field Theory for Any Brane Number
}}

\author{
Hiroyuki {\sc Hata}\footnote{
{\tt hata@gauge.scphys.kyoto-u.ac.jp}}
\\[7mm]
{\it
Department of Physics, Kyoto University, Kyoto 606-8502, Japan
}
}

\date{{\normalsize January 2019}}
\maketitle

\begin{abstract}
\normalsize

We present an analytic construction of multi-brane solutions with any
integer brane number in cubic open string field theory (CSFT) on the
basis of the $\KBc$ algebra.
Our solution is given in the pure-gauge form $\Psi=U\QB U^{-1}$ by
a unitary string field $U$, which we choose to satisfy two
requirements. First, the energy density of the solution should
reproduce that of the $(N+1)$-branes. Second, the EOM of the solution
should hold against the solution itself.
In spite of the pure-gauge form of $\Psi$, these two conditions are
non-trivial ones due to the singularity at $K=0$.
For the $(N+1)$-brane solution, our $U$ is specified by $[N/2]$
independent real parameters $\alpha_k$.
For the 2-brane ($N=1$), the solution is unique and reproduces the
known one.
We find that $\alpha_k$ satisfying the two conditions indeed
exist as far as we have tested for various integer values of
$N$ $(=2, 3, 4, 5, \cdots)$.
Our multi-brane solutions consisting only of the elements of
the $\KBc$ algebra have the problem that the EOM is not satisfied
against the Fock states and therefore are not complete ones.
However, our construction should be an important step toward
understanding the topological nature of CSFT which has similarities to
the Chern-Simons theory in three dimensions.
\end{abstract}

\thispagestyle{empty}
\end{titlepage}

\section{Introduction}

Since Schnabl's construction \cite{Schnabl} of an analytic
solution for tachyon condensation in cubic open string field theory
(CSFT), there have appeared lots of studies on
the analytic construction of solutions representing multiple
D25-branes within the framework of the $\KBc$ algebra
\cite{Okawa}.\footnote{
  For a recent numerical approach toward the construction of
  multi-brane   solutions, see ref.\ \cite{KudrnaSchnabl}.
}
Among them, the construction presented in \cite{ErlerMacc} by using
the boundary condition changing operators,
in addition to the elements of the $\KBc$ algebra,
may be a satisfactory one.
However, in this paper, we pursue the construction of multi-brane
solutions consisting solely of $(K,B,c)$.
Such types of solutions have been studied, for example, in
\cite{MS1,HKwn,MS2,HKinfty}, where they considered candidate solutions
of the pure-gauge type $\Psi=U\QB U^{-1}$ given in terms of $U$ and
$U^{-1}$ of the following form \cite{Okawa}:
\begin{equation}
U=\left(1-\sqrt{1-G}Bc\sqrt{1-G}\right)\frac{1}{\sqrt{G}} ,
\qquad
U^{-1}=U^\dg
=\frac{1}{\sqrt{G}}\left(G+\sqrt{1-G}Bc\sqrt{1-G}\right) ,
\label{eq:unitaryU_ES}
\end{equation}
where $G=G(K)$ is a function of $K$ which should suitably be
chosen.
Explicitly, $\Psi$ reads
\begin{equation}
\Psi=U\QB U^{-1}=\sqrt{1-G}\,cK\frac{1}{G}Bc\sqrt{1-G} .
\label{eq:Psi_ES}
\end{equation}
The $\KBc$ algebra we need here and in the following are
\begin{equation}
\CR{K}{B}=0,
\qquad
\ACR{B}{c}=1,
\qquad
B^2=c^2=0 ,
\label{eq:KBcAlgebra}
\end{equation}
and
\begin{equation}
\QB B=K,
\qquad
\QB K=0,
\qquad
\QB c=cKc .
\end{equation}
The string field $\Psi$ (as well as the elements $(K,B,c)$ of the
$\KBc$ algebra) is subject to the self-conjugateness condition
$\Psi^\dg=\Psi$ with $\dg$ denoting the composition of the
BPZ and the hermitian conjugations.
Therefore, $U$ in \eqref{eq:unitaryU_ES} is chosen to be unitary in
the sense that $U^\dg=U^{-1}$.
In fact, $U$ in \eqref{eq:unitaryU_ES} is the most generic form
of unitary $U$ which is the sum of two terms; one containing $Bc$ and
the other without it.

Though the configuration \eqref{eq:Psi_ES} is a pure-gauge one and
formally satisfies the EOM,
\begin{equation}
\QB\Psi+\Psi^2=0 ,
\label{eq:EOM}
\end{equation}
this is in fact a subtle problem due to the singularity at $K=0$.
As the requirements on the pure-gauge configuration $\Psi$
\eqref{eq:Psi_ES} as a solution,
the number of D25-branes $\Psi$ represents and the EOM test of $\Psi$
against itself were examined for various $G(K)$ defining $U$
\cite{MS1,HKwn,MS2,HKinfty}.
For calculating these quantities, we have to regularize the singularity
at $K=0$. In \cite{HKwn}, we adopted the $\Ke$-regularization of
replacing $K$ by $\Ke=K+\veps$ with $\veps$ being an infinitesimal
positive constant.
In this paper, for any $\calO(K,B,c)$, $\calO_\veps$ with subscript
$\veps$ denotes the $\Ke$-regularized one:
\begin{equation}
\calO_\veps=\calO\bigr|_{K\mapsto\Ke}=\calO(\Ke,B,c) .
\label{eq:calO_veps}
\end{equation}
Then, in terms of the $\Ke$-regularized pure-gauge configuration,
\begin{equation}
\Psie=\left(U\QB U^{-1}\right)_\veps ,
\label{eq:Psie}
\end{equation}
the brane number is given by $\calN+1$ with $\calN$ being
\begin{equation}
\calN=\frac{\pi^2}{3}\int\!\Psie^3 ,
\label{eq:calN}
\end{equation}
while the EOM test $\calT$ of $\Psie$ against itself is
\begin{equation}
\calT=\int\!\Psie\left(\QB\Psie+\Psie^2\right) .
\label{eq:calT}
\end{equation}
In fact, $\calN$ \eqref{eq:calN} is equal to the minus of the action
of $\Psie$,
$-S=-\int\left(\frac12\Psie\QB\Psie+\frac13\Psie^3\right)$,
divided by the D25-brane tension $1/(2\pi^2)$ only when the EOM test
\eqref{eq:calT} vanishes, $\calT=0$.\footnote{
  We are taking both the open string coupling constant and the
  space-time volume equal to one.}

The tachyon vacuum with $\calN=-1$ and the 2-brane with $\calN=1$ are
realized by \eqref{eq:Psi_ES} by taking $G(K)$ with its small $K$
behavior given by $G(K)\sim K$ and $G(K)\sim 1/K$,
respectively.\footnote{
  $G(K)$ should not have zero nor pole at $K=\infty$ to avoid their
  additional contribution to $\calN$ \cite{HKinfty}.
}
Concrete choices for $G(K)$ are, for example \cite{ES09,MS1,HKwn,MS2},
\begin{equation}
G_\textrm{tachyon vac.}(K)=\frac{K}{1+K},
\qquad
G_\textrm{2-brane}(K)=\frac{1+K}{K} .
\label{eq:G_tv_2-brane}
\end{equation}
The EOM test is also passed, namely, $\calT=0$ in these two cases.
It was shown that the origin of non-trivial $\calN$ in these solutions
is the singularity coming from the zero or pole of $G(K)$ at $K=0$
\cite{MS1,HKwn,MS2,HKinfty}.

However, the construction of multi-brane solutions with a larger
$\calN$ has been problematic.
From the above two examples in \eqref{eq:G_tv_2-brane}, it may be
guessed that a solution with $\calN=N=2,3,\cdots$ is obtained by
taking $G(K)$ with a multiple pole at $K=0$,
$G(K)\sim 1/K^N$ ($K\sim 0$);
for example, $G(K)=\left((1+K)/K\right)^N$.
However, it was found that $\calN$ and $\calT$ for this type of $G(K)$
are given by \cite{HKwn,MS2,HKinfty}
\begin{equation}
\calN=N+A_N,
\qquad
\calT=B_N ,
\label{eq:calN=N+A_N,calT=B_N}
\end{equation}
where the ``anomalous terms'' $A_N$ and $B_N$ are expressed in terms
of the confluent hypergeometric function\footnote{
  The confluent hypergeometric function is defined by
  $$
  {}_1F_1(a,b;z)=1+\sum_{k=1}^\infty\frac{a(a+1)\cdots (a+k-1)}
  {b(b+1)\cdots(b+k-1)}\frac{z^k}{k!} .
  $$
  Note that ${}_1F_1(a,b;z)$ is a polynomial in $z$ of degree $(-a)$
  for a non-positive integer $a$.
}
as
\begin{align}
A_N&=-\frac{\pi^2}{3}\,N\left(N^2-1\right)
\Re\,{}_1F_1(2-N,4;2\pi i) ,
\nn\\
B_N&=\frac{N(N+1)}{\pi}\,\Im\,{}_1F_1(1-N,2;2\pi i) .
\label{eq:A_N_B_N}
\end{align}
Examples are as follows:
\begin{equation}
A_N=
\begin{cases}
0 & (N=0,\pm 1) \\
-2\pi^2 &(N=2) \\
-8\pi^2 & (N=3) \\
-20\pi^2+4\pi^4 & (N=4)
\end{cases},
\qquad\quad
B_N=
\begin{cases}
0 & (N=0,\pm 1) \\
-6 &(N=2) \\
-24 & (N=3) \\
-60+(20/3)\pi^2 & (N=4)
\end{cases}.
\label{eq:A_N,B_N}
\end{equation}
Namely, $\calN$ is not an integer and the EOM test is not passed
($\calT\ne 0$) for the present type of solutions with $N\ge 2$.

In \cite{HKinfty}, we proposed that the $3$-brane solution with
$\calN=2$ and $\calT=0$ can be constructed in the form
\eqref{eq:Psi_ES} by making use of the singularities both at $K=0$ and
$K=\infty$, and taking, for example, $G(K)=(1+K)^2/K$.
However, multi-brane solutions with larger $\calN$ $(=3,4,5,\cdots)$
and $\calT=0$ seem not to exist in the form of \eqref{eq:Psi_ES}.

In this paper, we present an analytic expression of
multi-brane solutions carrying any integer $\calN$ and satisfying the
EOM test $\calT=0$.
We start with the most generic form of unitary string field
$U$ consisting only of $(K,B,c)$ and examine the pure-gauge
configuration $\Psi=U\QB U^{-1}$ which manifestly satisfies the
self-conjugateness condition.
For considering the most generic unitary $U$, we adopt a convenient
notation for expressing a string field which is given as the
sum of products of $(K,B,c)$.
Then, by referring to the successful examples
of the tachyon vacuum and the $2$-brane solutions given by
\eqref{eq:unitaryU_ES}, \eqref{eq:Psi_ES} and \eqref{eq:G_tv_2-brane},
we make a natural ansatz on the functions of $K$ defining $U$. As a
result, $U$ which is expected to represent $(N+1)$-branes is specified
by $(N+1)$ real parameters
$\left(\alpha_0,\alpha_1,\cdots,\alpha_N\right)$,
among which only $[N/2]$ are independent.\footnote{
  $[x]$ denotes the greatest integer less than or equal to $x$.
}
We carry out the calculation of $\calN$ \eqref{eq:calN} and $\calT$
\eqref{eq:calT} for this type of solution, and find that
these two quantities are again given in the form
\eqref{eq:calN=N+A_N,calT=B_N}: the anomalous terms $A_N$ and
$B_N$ are polynomials in $(2\pi i)^2$ of order $[N/2]$ and $[N/2]-1$,
respectively ($A_N$ starts with the $(2\pi i)^2$ term). This is also
the case for $A_N$ and $B_N$ of \eqref{eq:A_N_B_N} for the solution
\eqref{eq:Psi_ES}.
A different point in the present $U$ is that the
coefficients of the polynomials are not constants but are linear
functions of $\alpha_k$.
Moreover, the coefficient $f_n(\alpha_k)$ multiplying $(2\pi i)^{2n}$
is common between $A_N$ and $(2\pi i)^2 B_N$ up to a constant factor.
Therefore, both $A_N=0$ and $B_N=0$, namely, $\calN=N$ and $\calT=0$,
are realized by choosing as $\{\alpha_k\}$ the solution to
$f_n(\alpha_k)=0$ ($n=1,2,\cdots,[N/2]$).
In fact, we find that $\alpha_k$ and hence the solution
$\Psi=U\QB U^{-1}$ are uniquely determined in this way for any
integer values of $N$ ($=2,3,4,5,\cdots$) we have tested.
For example, the $3$-brane solution is given by
\eqref{eq:Psi_3-brane} with $G=(1+K)/K$.

However, we have not succeeded in determining $\alpha_k$ for a generic
$N$. The reason is that the expressions of $\calN$ and $\calT$ we will
obtain in this paper are too complicated to get $f_n(\alpha_k)$ in a
closed form for a generic $N$. Even more, the fact that
$f_n(\alpha_k)$ are common between $A_N$ and $(2\pi i)^2B_N$ is merely
an ``experimental fact'' obtained by the evaluation of $A_N$ and $B_N$
for various values of $N$.
However, there is no doubt that we can determine $\alpha_k$ so that
our solution can realize both $\calN=N$ and $\calT=0$ for any integer
$N$.
The technical problem of giving $f_n(\alpha_k)$ for a generic $N$
will be resolved by mathematical sophistication.

Even if the solution $\{\alpha_k\}$ to $f_n(\alpha_k)=0$ is found for
a generic integer $N$, there still is an important problem in our
construction of solutions.
In this paper, as the EOM test, we consider only $\calT$
\eqref{eq:calT}, namely, the EOM test against the candidate solution
$\Psi$ itself.
However, it has been known that the 2-brane solution given by $U$ of
\eqref{eq:unitaryU_ES} with $G=G_\textrm{2-brane}$
\eqref{eq:G_tv_2-brane} does not pass the EOM test against the Fock
states \cite{MS2}, and this property is inherited by the multi-brane
solutions in this paper consisting solely of $(K,B,c)$
(see Sec.\ \ref{sec:summary}).
This problem of the failure of the EOM test against the Fock states
might be resolved by some improvements of the solution, or by some
consistent truncation of the space of fluctuations around
multi-branes which excludes the Fock states.

However, even if this problem persists, the construction in this paper
should give an important hint on understanding the meaning of 
$\calN$ \eqref{eq:calN} as ``winding number''.
Namely, note the analogy of $\calN$ \eqref{eq:calN} to the winding
number,
\begin{equation}
\calW[g]=\frac{1}{24\pi^2}\int_M\tr\left(gd g^{-1}\right)^3 ,
\label{eq:calW}
\end{equation}
of the mapping $g(x)$ from a three-manifold $M$ to a Lie group.
This analogy was emphasized and examined in \cite{HKwn}.
There, $\calN$ was evaluated by making use of its topological nature,
namely, the invariance of $\calN$ under small deformations of $U$,
to identify the zero or pole of $G(K)$ at $K=0$ as the origin of
non-trivial $\calN$ (see Sec.\ \ref{sec:calN} of this paper).
For explaining the relevance of the present construction of $U$ giving
integer $\calN$ to the identification of $\calN$ as winding number,
let us consider the simplest example of $\calW$;
$g(x)\in SU(2)$, $M=S^3$ and the hedgehog type
$g(x)=\exp\left(i f(r)\,\bm{x}\cdot\bm{\tau}/r\right)$
with $r=\abs{\bm{x}}$.
In this case, $\calW$ is given in terms of $f(r)$ at the origin and
the infinity by
$\calW=\left(f(\infty)-f(0)\right)/\pi$.
This $\calW$ becomes an integer by demanding the regularity of $g(x)$
at the two points, which implies that both $f(0)$ and $f(\infty)$ are
integer multiples of $\pi$.
The non-integer results \eqref{eq:calN=N+A_N,calT=B_N} and
\eqref{eq:A_N,B_N} of $\calN$ for $U$ of the form
\eqref{eq:unitaryU_ES} and our finding in this paper of new type of
$U$ realizing integer $\calN$ for larger $N$ may give a clue to
understanding the meaning of regularity of $U$.\footnote{
Naively, it is guessed that the two points $r=0, \infty$ in the
example of hedgehog $g(x)$ correspond to $K=0, \infty$ in CSFT.}
Of course, we have to find answers to more basic questions;
``What are the counterparts of the three-manifold
$M$ and the Lie group in CSFT? What is the meaning of winding
represented by $\calN$?''.
These considerations are further expected to lead to deeper
understanding of the similarity of CSFT to the Chern-Simons
theory in three dimensions, and topological aspects of CSFT.

The rest of this paper is organized as follows.
In Sec.\ 2, first introducing our convenient notation for expressing
the sum of products of $(K,B,c)$, we determine the form of the most
generic unitary string field $U$, and present our assumption on the
form of $U$ which is specified by $\alpha_k$.
Then, in Secs.\ 3 and 4, we obtain $\calN$ and $\calT$,
respectively, as functions of $\alpha_k$. In particular, we calculate
$\calN$ not directly but in a way where the role of the singularity
at $K=0$ as the origin of non-trivial $\calN$ is manifest.
In Sec.\ 5, we examine the conditions $\calN=N$ and $\calT=0$ on our
solution to determine $\alpha_k$ for various values of $N$.
We summarize the paper and discuss future problems in Sec.\ 6.
In the Appendices, we present technical details used in the text.

\section{Assumptions on the solution}
\label{sec:Assump_on_Sol}

In this section, we first introduce our convenient notation for
expressing string fields in the framework of the $\KBc$ algebra.
Then, we obtain the form of the most generic unitary string field $U$
for our candidate solution $\Psi=U\QB U^{-1}$ of the pure-gauge type.
After these preparations, we restrict $U$ to a particular form which
is specified by real parameters
$(\alpha_0,\alpha_1,\cdots,\alpha_N)$.

\subsection{Convenient notation}
\label{sec:ConvNotation}

For making our equations look simpler, we first introduce a convenient
notation for expressing the sum of products of $(K,B,c)$.
Let us consider, for example, the following string field $\calO$:
\begin{equation}
\calO=\sum_{\{f_a\}}f_1(K)\,c\,f_2(K)\,c
\cdots c\,f_{n}(K)\,Bc\,f_{n+1}(K) ,
\label{eq:calO}
\end{equation}
where $f_a(K)$ ($a=1,2,\cdots,n+1$) are functions of $K$ and the sum
$\sum_{\{f_a\}}$ is the sum over various sets of $f_a$.
In this particular example, there appear $(n-1)$ ghosts $c$ and a
single $Bc$. Our new notation also applies to the cases where some of
the $c$'s are replaced with $Bc$ (and $Bc$ with $c$).

In our new notation, we first consider the product of $c$'s
only in \eqref{eq:calO} to write it as\linebreak
$c_{12}c_{23}\cdots c_{n-1,n}c_{n,n+1}$
by attaching to each $c$ a pair of numbers $(a,b)$.
Namely, each of the numbers $(1,2,\cdots,n,n+1)$ specifies a position in
the sequence of $c$'s.
Then, we assign each of $K$ and $B$ in \eqref{eq:calO} (which are
commutative with each other) a single number $a$ specifying their
position in the product of $c$'s to write $K_a$ and $B_a$.
Then, the string field $\calO$ \eqref{eq:calO} now carries a pair of
indices $(1,n+1)$ and is written as
\begin{equation}
\calO_{1,n+1}
=A_{1,2,\cdots,n,n+1}\,c_{12}\,c_{23}\,\cdots\,c_{n-1,n}(Bc)_{n,n+1} ,
\label{eq:calO_1,n+1}
\end{equation}
where $A_{1,2\cdots,n,n+1}$, which depends only on $K$, is given by
\begin{equation}
A_{1,2,\cdots,n,n+1}=\sum_{\{f_a\}}f_1(K_1)f_2(K_2)\cdots
f_n(K_n)f_{n+1}(K_{n+1}) .
\end{equation}
In \eqref{eq:calO_1,n+1} and in the following, we use notations
such as $(Bc)_{ab}\left(=B_ac_{ab}\right)$,
$(cB)_{ab}\left(=c_{ab}B_b\right)$ and $(cK)_{ab}\,(=c_{ab}K_b)$.
The advantage of the present notation is that we can put the
$K$-dependencies at any place without any ambiguity.

As examples, $U$ in \eqref{eq:unitaryU_ES} and $\Psi$
\eqref{eq:Psi_ES} are expressed in our notation as
\begin{equation}
U_{12}=\frac{1}{\sqrt{G_2}}\id_{12}
-\frac{\sqrt{\left(1-G_1\right)\left(1-G_2\right)}}{\sqrt{G_2}}
(Bc)_{12} ,
\label{eq:U_12_N=1}
\end{equation}
and
\begin{equation}
\Psi_{13}=\sqrt{1-G_1}\,\frac{K_2}{G_2}\sqrt{1-G_3}
\,c_{12}(Bc)_{23} ,
\end{equation}
with $G_a=G(K_a)$ and $\id_{ab}$ being the identity string field.
Finally, the conjugate of $\calO$ \eqref{eq:calO_1,n+1} for a
self-conjugate $A_{1,2,\cdots,n,n+1}$ is given by
\begin{equation}
(\calO^\dg)_{n+1,1}=A_{1,2,\cdots,n,n+1}
\,(cB)_{n+1,n}\,c_{n,n-1}\,\cdots c_{32}\,c_{21} .
\label{eq:calO^dagger}
\end{equation}

\subsection{The most generic unitary $U$}
\label{sec:MostGenUnitaryU}

For constructing self-conjugate solutions in the pure-gauge form,
$\Psi=U\QB U^{-1}$, in terms of a unitary $U$ satisfying
$UU^\dg=\id$,
let us first establish the most general form of the string field $U$
which is unitary and carries the ghost-number $\Ngh=0$.
First, from $\Ngh=0$, $U$ is expressed without losing generality as
\begin{equation}
U_{12}=\frac{1}{\Gm_2}\id_{12}-\frac{\pF_{12}}{\Gm_2}(Bc)_{12} ,
\label{eq:U_by_Gm_pF}
\end{equation}
where $\Gm_2$ and $\pF_{12}$ on the RHS are given by $\Gm_a=\Gm(K_a)$
and $\pF_{ab}=\pF(K_a,K_b)$ in terms of two real functions $\Gm(x)$
and $\pF(x,y)$.\footnote{
  We are assuming that $U$ is real, namely, that $U$ does not contain
  any imaginary unit $i$. This reality assumption is only for the
  sake of simplicity.}
Then, $U$ is unitary if $\pF_{aa}$ and $\Gm_a$ are related by
\begin{equation}
\pF_{aa}=1-(\Gm_a)^2 ,
\label{eq:pF_aa=1-Gm_a^2}
\end{equation}
and $\pF_{ab}$ is symmetric:
\begin{equation}
\pF_{ab}=\pF_{ba} .
\label{eq:pF_ab=pF_ba}
\end{equation}
The derivation of these two conditions as well as
those of some of the equations in this subsection are given in
Appendix \ref{app:MostGenUnitaryU}.

When the two conditions \eqref{eq:pF_aa=1-Gm_a^2} and
\eqref{eq:pF_ab=pF_ba} are met, $U^{-1}$ is given by
\begin{equation}
\bigl(U^{-1}\bigr)_{12}=\bigl(U^\dg\bigr)_{12}
=\Gm_1\id_{12}+\frac{\pF_{12}}{\Gm_1}(Bc)_{12} ,
\label{eq:U^-1=U^dagger=}
\end{equation}
and the corresponding candidate solution of the pure-gauge type 
$U\QB U^{-1}$ reads
\begin{equation}
\left(U\QB U^{-1}\right)_{13}
=E_{123}\left(cK\right)_{12}\left(Bc\right)_{23} ,
\label{eq:UQBU^-1}
\end{equation}
where $E_{abc}$ is defined by
\begin{equation}
E_{abc}=\pF_{ac}+\pF_{ab}\,\frac{1}{\Gm_b^2}\,\pF_{bc} .
\label{eq:E_abc}
\end{equation}
We summarize three kinds of relations concerning $E_{abc}$
\eqref{eq:E_abc}:
\begin{align}
&E_{abc}=E_{cba} ,
\\
&E_{aab}=\frac{1}{\Gm_a^2}\,\pF_{ab} ,
\qquad
E_{abb}=\pF_{ab}\,\frac{1}{\Gm_b^2} ,
\qquad
E_{aaa}=\frac{1}{\Gm_a^2}-1 ,
\\
&E_{abb}E_{bcd}-E_{abc}E_{ccd}=E_{abd}-E_{acd} .
\label{eq:EE-EE=E-E}
\end{align}

\subsection{Assumptions on $\Gm_a$ and $\pF_{ab}$}

It is impossible to evaluate $\calN$ and $\calT$ for
$\Psi=U\QB U^{-1}$ given by \eqref{eq:UQBU^-1} without any
assumptions on $\Gm_a$ and $\pF_{ab}$.
Here, on the basis of known facts, we would like to 
make plausible assumptions on the form of $\Gm_a$ and $\pF_{ab}$
which is expected to realize $\calN=N$ and $\calT=0$ for each positive
integer $N$.

The first fact is the satisfactory example of the $\calN=1$ solution
with $U$ given by \eqref{eq:unitaryU_ES} or by \eqref{eq:U_12_N=1} in
the present notation. In this case, $G(K)$ should have a simple pole
at $K=0$ and no other zeros/poles in the complex half-plane
$\Re K\ge 0$ including $K=\infty$,
but otherwise arbitrary. For definiteness, we take
\begin{equation}
G(K)=\frac{1+K}{K} .
\label{eq:G_fixed}
\end{equation}
Comparing \eqref{eq:U_12_N=1} with the generic form
\eqref{eq:U_by_Gm_pF}, we see that $\Gm_a$ and $\pF_{ab}$ in this
example are given by
\begin{equation}
\Gm_a=\sqrt{G_a}=\sqrt{G(K_a)},
\qquad
\pF_{ab}=\sqrt{\left(1-G_a\right)\left(1-G_b\right)} ,
\label{eq:Gm_pF_N=1}
\end{equation}
which certainly satisfy \eqref{eq:pF_aa=1-Gm_a^2} and
\eqref{eq:pF_ab=pF_ba}.

Secondly, by replacing $G_a$ in \eqref{eq:Gm_pF_N=1} with
$G_a^N=\left((1+K_a)/K_a\right)^N$, we get $\calN$ and $\calT$ given
by \eqref{eq:calN=N+A_N,calT=B_N} and \eqref{eq:A_N_B_N}.
As we saw there, $\calN$ for $N\ge 2$ is a polynomial
in $(2\pi i)^2$ starting with the zero-th term $N$.
This seems to suggest that the replacement $G_a\mapsto (G_a)^N$ is,
though not perfect, fairly close to the final answer realizing
$\calN=N$.

Taking these facts into account, let us take as our candidate $\Gm_a$
and $\pF_{ab}$ for a generic $N$, which possibly realize $\calN=N$ and
$\calT=0$, the following ones given in terms of $G(K)$ of
\eqref{eq:G_fixed}:
\begin{align}
\Gm_a&=G_a^{N/2}=G(K_a)^{N/2} ,
\label{eq:Gm_a=}
\\
\pF_{ab}&=\prod_{k=0}^N\left(1-G_a^k\,G_b^{N-k}\right)^{\alpha_k}
=-\prod_{k=0}^N\left(G_a^k\,G_b^{N-k}-1\right)^{\alpha_k} .
\label{eq:pF_ab=}
\end{align}
Here, $\alpha_k$ are numerical coefficients satisfying
\begin{equation}
\sum_{k=0}^N\alpha_k=1,
\label{eq:sum_kalpha_k=1}
\end{equation}
and
\begin{equation}
\alpha_{N-k}=\alpha_k
\qquad\left(k=0,1,\cdots,N\right) .
\label{eq:alpha_k=alpha_N-k}
\end{equation}
These conditions \eqref{eq:sum_kalpha_k=1} and
\eqref{eq:alpha_k=alpha_N-k} are necessary for
\eqref{eq:pF_aa=1-Gm_a^2} and \eqref{eq:pF_ab=pF_ba}, respectively.
Note that
\begin{equation}
\sum_{k=0}^Nk\alpha_k=\frac{N}{2} ,
\label{eq:sum_kkalpha_k=N/2}
\end{equation}
follows from \eqref{eq:sum_kalpha_k=1} and
\eqref{eq:alpha_k=alpha_N-k}.
The simple replacement $G_a\mapsto(G_a)^N$ in \eqref{eq:Gm_pF_N=1}
corresponds to the following choice of $\alpha_k$:
\begin{equation}
\alpha_0=\alpha_N=\frac12,
\qquad \mbox{other }\alpha_k=0 .
\end{equation}

Though $\pF_{ab}$ \eqref{eq:pF_ab=} itself is not of a factorized form
with respect to the $K_a$ and $K_b$ dependences, it should be
suitably expressed as a sum of factorized terms by, for example,
Taylor expansion, for calculating correlators containing $\pF_{ab}$.

In the rest of this paper, we shall first obtain $\calN$ and $\calT$
for the present solution as functions of $\{\alpha_k\}$, and then
examine whether there exists $\{\alpha_k\}$ satisfying both
$\calN[\alpha_k]=N$ and $\calT[\alpha_k]=0$ for each positive integer
$N$.

\section{Expression of $\calN[\alpha_k]$}
\label{sec:calN}

As a preparation for examining $\calN$ \eqref{eq:calN} for our
candidate solution proposed above, we in this section obtain a
calculable concrete expression of $\calN[\alpha_k]$ for a given
$\{\alpha_k\}$.

\subsection{$\calN$ in terms of $\Gm_a$ and $\pF_{ab}$}
\label{sec:calN_by_Gma_pFab}

Instead of calculating $\calN$ \eqref{eq:calN} directly, we here use
the method of ref.\ \cite{HKwn} to evaluate $\calN$ as a
``topological'' quantity. Concretely, we use the following formula
for the variation of $\calN$ under an arbitrary infinitesimal
deformation $\delta U^{-1}$ of $U^{-1}$:
\begin{equation}
\frac{1}{\pi^2}\,\delta\calN
=\veps\int\!\left\{T_B\!\left[\left(U\QB U^{-1}\right)^2
\left(U\,\delta U^{-1}\right)\right]\right\}_\veps .
\label{eq:delta_calN=}
\end{equation}
Here, $T_B$ is the Grassmann-odd operation of replacing $B$ with
the identity $\id$ one by one:
\begin{align}
T_B\!\left(f_1 B f_2 B f_3\cdots f_n B f_{n+1}\right)
&=(-1)^{\abs{f_1}}f_1\id f_2 B f_3\cdots f_nBf_{n+1}
+(-1)^{\abs{f_1}+\abs{f_2}+1}f_1 B f_2\id f_3\cdots f_n Bf_{n+1}
\nn\\
&\qquad
+\ldots +(-1)^{\sum_{i=1}^n\abs{f_i}+n-1}
f_1 B f_2 B f_3\cdots f_n \id f_{n+1} ,
\label{eq:T_B}
\end{align}
where $f_i=f_i(c,K)$ is a product of $K$'s and $c$'s, and $\abs{f}=0$
($=1$) if $f$ is Grassmann-even (-odd). The operation of $T_B$
on a quantity without $B$ is defined to be zero:
\begin{equation}
T_B f(c,K)=0 .
\end{equation}
The derivation of \eqref{eq:delta_calN=} is given in Appendix
\ref{app:Deriv_deltacalN}.

For calculating $\calN$ of our solution with $G(K)$ given by
\eqref{eq:G_fixed}, we introduce $G(K,u)$ with a parameter $u$,
\begin{equation}
G(K,u)=\frac{1+K}{u+K} ,
\label{eq:G(K,u)}
\end{equation}
and regard the deformation $\delta$ as that of $u$:
$\delta=\delta u\left(d/du\right)$.
Then, since $G(K,u)$ with $u>0$ corresponds to the trivial solution
with $\calN=0$, $\calN$ for $G(K)$ \eqref{eq:G_fixed} is given by
integrating \eqref{eq:delta_calN=} over $u$ as
\begin{equation}
\calN=\int_{u=u_0}^{u=0}\!\delta\calN ,
\label{eq:calN_by_u-int}
\end{equation}
where $\delta\calN$ on the RHS is that for $G(K,u)$ \eqref{eq:G(K,u)},
and $u_0$ is positive but otherwise arbitrary (the integration
\eqref{eq:calN_by_u-int} is independent of $u_0$ in the limit
$\veps\to +0$).
Eq.\ \eqref{eq:calN_by_u-int} which is multiplied by $\veps$ can be
non-vanishing in the limit $\veps\to +0$ due to the $1/\veps$
singularity arising from the $u$-integration near $u=0$ as we saw in
\cite{HKwn}.

Let us express the integrand on the RHS of \eqref{eq:delta_calN=} in
terms of $\Gm_a$ and $\pF_{ab}$.
The expression of $U\QB U^{-1}$ is already given by \eqref{eq:UQBU^-1}
and \eqref{eq:E_abc}.
Using this, $\left(U\QB U^{-1}\right)^2$ is calculated as follows:
\begin{align}
\bigl[(U\QB U^{-1})^2\bigr]_{15}
&=\left(U\QB U^{-1}\right)_{13}\left(U\QB U^{-1}\right)_{35}
=E_{123}E_{345}\left(cK\right)_{12}\left(Bc\right)_{23}
\left(cK\right)_{34}\left(Bc\right)_{45}
\nn\\
&=\left(E_{122}E_{235}-E_{123}E_{335}\right)
\left(cK\right)_{12}\left(cK\right)_{23}\left(Bc\right)_{35}
\nn\\
&=\left(E_{125}-E_{135}\right)
\left(cK\right)_{12}\left(cK\right)_{23}\left(Bc\right)_{35} ,
\label{eq:(UQBU^-1)^2}
\end{align}
where we have used \eqref{eq:Bc_cK_Bc=},
and the last equality is due to the relation \eqref{eq:EE-EE=E-E}.
Next, let us consider $U\delta U^{-1}$.
Taking the variation of $U^{-1}$ \eqref{eq:U^-1=U^dagger=}, we obtain
\begin{equation}
\left(\delta U^{-1}\right)_{12}
=\Gm_1\left(\delta\ln \Gm_1\right)\id_{12}
+\frac{1}{\Gm_1}\bigl(\delta\pF_{12}
-\left(\delta\ln\Gm_1\right)\pF_{12}\bigr)(Bc)_{12} .
\end{equation}
Using this, $U\delta U^{-1}$ is calculated as follows:
\begin{align}
\left(U\delta U^{-1}\right)_{13}
&=U_{12}\left(\delta U^{-1}\right)_{23}
\nn\\
&=\left(\frac{1}{\Gm_2}\id_{12}
-\frac{\pF_{12}}{\Gm_2}(Bc)_{12}\right)
\left[\Gm_2\left(\delta\ln \Gm_2\right)\id_{23}
+\frac{1}{\Gm_2}\bigl(\delta\pF_{23}
-\left(\delta\ln\Gm_2\right)\pF_{23}\bigr)(Bc)_{23}\right]
\nn\\
&=\left(\delta\ln \Gm_1\right)\id_{13}
+\bigl[\delta\pF_{13}-\left(\delta\ln\Gm_1+\delta\ln\Gm_3\right)
\pF_{13}\bigr](Bc)_{13} ,
\label{eq:UdeltaU^-1}
\end{align}
where we have used \eqref{eq:Bc_Bc=} and \eqref{eq:pF_aa=1-Gm_a^2}.
Finally, multiplying \eqref{eq:(UQBU^-1)^2} and \eqref{eq:UdeltaU^-1},
we get
\begin{align}
&\left[(U\QB U^{-1})^2 U\delta U^{-1}\right]_{15}
=\bigl[(U\QB U^{-1})^2\bigr]_{14}\left(U\delta U^{-1}\right)_{45}
\nn\\
&=\Bigl\{\left(E_{125}-E_{135}\right)\delta\ln\Gm_5
+\left(E_{123}-E_{133}\right)
\bigl[\delta\pF_{35}-\left(\delta\ln\Gm_3+\delta\ln\Gm_5\right)
\pF_{35}\bigr]\Bigr\}
\nn\\
&\qquad\times
\left(cK\right)_{12}\left(cK\right)_{23}\left(Bc\right)_{35} .
\label{eq:(UQBU^-1)^2(UdeltaU^-1)}
\end{align}
By the substitution of this into \eqref{eq:delta_calN=}, $(Bc)_{35}$
is replaced with $c_{35}$ by the $T_B$ operation, and the last index
$5$ is identified with the first index $1$. Then, we get the desired
formula for calculating $\calN$:
\begin{equation}
\frac{1}{\pi^2}\,\delta\calN
=\veps\int\!\left(W_{123}\right)_\veps
\left(c\Ke\right)_{12}\left(c\Ke\right)_{23}c_{31} ,
\label{eq:deltacalN_by_W}
\end{equation}
with $W_{123}$ given by
\begin{equation}
W_{123}=\left(E_{123}-E_{133}\right)\bigl[\delta\pF_{31}
-\left(\delta\ln\Gm_3+\delta\ln\Gm_1\right)\pF_{31}\bigr] .
\label{eq:W=}
\end{equation}
Note that the
$\left(E_{125}-E_{135}\right)\delta\ln\Gm_5$ term
in \eqref{eq:(UQBU^-1)^2(UdeltaU^-1)} does not contribute to
\eqref{eq:deltacalN_by_W} due to the L/R-reversing symmetry of the
$ccc$-correlator:
\begin{equation}
\int\!A_{123}\,c_{12}\,c_{23}\,c_{31}
=\int\!A_{132}\,c_{12}\,c_{23}\,c_{31} ,
\label{eq:L/R-sym}
\end{equation}
valid for any $A_{123}(K)$.

Though we do not use it in this paper, $\calN$ itself is of course
given in terms of $E_{abc}$:
\begin{equation}
\frac{1}{\pi^2}\,\calN=\int\!\left(M_{1234}\right)_\veps
\left(c\Ke\right)_{12}\left(c\Ke\right)_{23}\left(c\Ke\right)_{34}
\left(Bc\right)_{41} ,
\label{eq:calN=}
\end{equation}
where $M_{1234}$ is
\begin{equation}
M_{1234}=\left(E_{123}-E_{133}\right)E_{341}
-\left(E_{124}-E_{134}\right)E_{441} .
\label{eq:M_1234}
\end{equation}

\subsection{$\calN$ for a given  $\{\alpha_k\}$}
\label{sec:calN[alpha_k]}

The formula \eqref{eq:deltacalN_by_W} is valid for any $U$
\eqref{eq:U_by_Gm_pF} given in terms of
$\left(\Gm_a, \pF_{ab}\right)$.
In this subsection, we use \eqref{eq:deltacalN_by_W} and
\eqref{eq:calN_by_u-int} to calculate 
$\calN$ for our particular choice of $\Gm_a$ and $\pF_{ab}$,
\eqref{eq:Gm_a=} and \eqref{eq:pF_ab=}, specified by $\{\alpha_k\}$.
An important point in this calculation is that
\eqref{eq:deltacalN_by_W} is multiplied by $\veps$ which should be
taken to $+0$ in the end.
This implies that we are allowed to keep only the most singular part
of $W_{123}$ \eqref{eq:W=} with respect to $\veps$.

Recall that $G_a$ in $W_{123}$ \eqref{eq:W=} is given by
\eqref{eq:G(K,u)} with the parameter $u$, and the variation $\delta$
is that with respect to $u$.
The $\Ke$-regularized $G_a$ in $(W_{123})_\veps$ is taken as
\begin{equation}
G(\Ke, u)=\frac{1}{u+\Ke} ,
\label{eq:G(Ke,u)}
\end{equation}
where $\Ke$ in the numerator of the original $G(\Ke,u)$
corresponding to \eqref{eq:G(K,u)} has been omitted since it is
irrelevant (i.e., higher order in $\veps$) in the present
calculation.
We regard this $G(\Ke, u)$ as an $O(1/\veps)$ quantity\footnote{
  $u$ can also be regarded as $O(\veps)$ since only the
  part $0\le u< O(\veps)$ of the $u$-integration region contributes to
  \eqref{eq:calN_by_u-int}.
}
and expand $W_{123}$ in inverse powers of $G_a$.
In the following calculations, the properties of $\alpha_k$,
\eqref{eq:sum_kalpha_k=1}, \eqref{eq:alpha_k=alpha_N-k} and
\eqref{eq:sum_kkalpha_k=N/2}, are repeatedly used without mentioning
it.
In addition, we omit the subscript $\veps$ in \eqref{eq:calO_veps}
for the replacement $K\mapsto\Ke$ for the sake of notational
simplicity.
Everything in this subsection should be regarded as the
$\Ke$-regularized one.

First, we obtain without approximation
\begin{equation}
\delta\pF_{ab}
-\left(\delta\ln\Gm_a+\delta\ln\Gm_b\right)\pF_{ab}
=\sum_{k=0}^N\alpha_k
\frac{k\,\delta\ln G_a+\left(N-k\right)\delta\ln G_b}{
  G_a^k\,G_b^{N-k}-1}\,\pF_{ab} ,
\label{eq:delta_pF_ab=}
\end{equation}
where we have used
$\delta\ln\Gm_a=\left(N/2\right)\delta\ln G_a$.
Next, $\pF_{ab}$ is expanded in inverse powers of $G_a$ as
\begin{equation}
\pF_{ab}=-\left(G_a G_b\right)^{N/2}\left[
1-\sum_{k=0}^N\frac{\alpha_k}{G_a^k\,G_b^{N-k}}
+O\!\left(\frac{1}{G^{2N}}\right)\right] ,
\label{eq:Expand_pF_ab}
\end{equation}
and, using this, $E_{abc}$ \eqref{eq:E_abc} is expanded as
\begin{equation}
E_{abc}=\left(G_a\,G_c\right)^{N/2}
\sum_{k=0}^N\alpha_k\left[\frac{1}{G_a^k\,G_c^{N-k}}
-\frac{1}{G_a^k\,G_b^{N-k}}-\frac{1}{G_b^k\,G_c^{N-k}}
\right]+O\!\left(\frac{1}{G^{N}}\right) .
\label{eq:Expand_E_abc}
\end{equation}
From \eqref{eq:delta_pF_ab=}, \eqref{eq:Expand_pF_ab} and
\eqref{eq:Expand_E_abc}, we obtain the following expansion of
$W_{123}$:
\begin{align}
W_{123}&=-\sum_{k=0}^N\alpha_k\left[
\left(\frac{G_3}{G_1}\right)^N\left(\frac{G_1}{G_2}\right)^k
+\left(\frac{G_3}{G_2}\right)^k
-\left(\frac{G_3}{G_1}\right)^k-1\right]
\nn\\
&\qquad\quad\times
\sum_{\ell=0}^N\alpha_\ell
\bigl[\ell\,G_3+\left(N-\ell\right)G_1\bigr]
\left(\frac{G_1}{G_3}\right)^\ell\times \delta u
+O\!\left(\frac{\delta\ln G}{G^N}\right) .
\label{eq:Expand_W_123}
\end{align}
In \eqref{eq:Expand_W_123}, we have used that $\delta\ln G$ for $G$ of
\eqref{eq:G(Ke,u)} is given by
\begin{equation}
\delta\ln G=-\frac{\delta u}{u+\Ke}=-G\,\delta u .
\label{eq:deltalnG=}
\end{equation}

As seen from \eqref{eq:Expand_W_123}, the leading part of $W_{123}$ is
the sum of terms of the form
$G_1^{n_1}\,G_2^{n_2}\,G_3^{n_3}\,\delta u$
with integers $n_a$ satisfying $n_1+n_2+n_3=1$.
As we shall see,
this leading part makes finite $O(\veps^0)$ contribution to $\calN$,
while the contribution of the subleading part is of $O(\veps^N)$.
Therefore, we keep only the leading part of $W_{123}$ in the rest of
this subsection. Then, defining $\calN_{n_1,n_2,n_3}$ by
\begin{equation}
\frac{1}{\pi^2}\,\calN_{n_1,n_2,n_3}
=\veps\int_{u_0}^0\!du\int\!G_1^{n_1}\,G_2^{n_2}\,G_3^{n_3}
\,\left(c\Ke\right)_{12}\left(c\Ke\right)_{23}c_{31}
\qquad
\biggl(\sum_{a=1}^3n_a=1\biggr) ,
\label{eq:calN_n1n2n3}
\end{equation}
we see that $\calN[\alpha_k]$ is given by
\begin{align}
\calN[\alpha_k]
&=-\sum_{k,\ell=0}^N\alpha_k\alpha_\ell\biggl[
\ell\Bigl(\calN_{k-\ell+1,-k,\ell}
+\calN_{\ell,-k,k-\ell+1}-\calN_{\ell-k,0,k-\ell+1}
-\calN_{\ell,0,-\ell+1}\Bigr)
\nn\\
&\qquad
+\left(N-\ell\right)\Bigl(
\calN_{k-\ell,-k,\ell+1}
+\calN_{\ell+1,-k,k-\ell}
-\calN_{\ell-k+1,0,k-\ell}
-\calN_{\ell+1,0,-\ell}\Bigr)\biggr] ,
\label{eq:calN_by_calN_n1n2n3}
\end{align}
where we have made the replacement $\ell\to N-\ell$ for a number of
terms to eliminate $N$ from their indices.

Next, using
\begin{equation}
\Ke=\frac{1}{G}-u ,
\end{equation}
for $c\Ke$ in \eqref{eq:calN_n1n2n3} and defining $S_{m_1,m_2,m_3}$ by
\begin{equation}
\frac{1}{\pi^2}S_{m_1,m_2,m_3}
=\veps\int_{u_0}^0\!du\,u^{1+\sum_{a=1}^3m_a}\,
\int\!c_{12}\,c_{23}\,c_{31}\,G_1^{m_1}G_2^{m_2}G_3^{m_3}
\qquad
\left(\sum_{a=1}^3 m_a=-1,0,1\right) ,
\label{eq:S_m1m2m3}
\end{equation}
$\calN_{n_1,n_2,n_3}$ \eqref{eq:calN_n1n2n3} is given as
\begin{equation}
\calN_{n_1,n_2,n_3}=S_{n_1,n_2,n_3}-S_{n_1,n_2-1,n_3}
-S_{n_1,n_2,n_3-1}+S_{n_1,n_2-1,n_3-1} .
\label{eq:calN_n1n2n3_by_S_m1m2m3}
\end{equation}
Note that $S_{m_1,m_2,m_3}$ is totally symmetric with respect to its
indices and vanishes if at least one of the three $m_a$ is equal to
zero.
We calculate $S_{m_1,m_2,m_3}$ \eqref{eq:S_m1m2m3} in
Appendix \ref{app:S_by_F} by using the $(s,z)$-integration method
of \cite{MS1,MS2}.\footnote{
The $(s,z)$-integration method has an ambiguity when the poles of the
$z$-integration are located on the imaginary axis ($\Re z=0$).
This ambiguity is avoided in the present case due to the
$\Ke$-regularization.
}
The results are as follows.
First, we introduce a function $F_{P,Q}(z)$ defined by a pair of
integers $(P,Q)$ with $P+Q=0,1$ or $2$:
\begin{align}
F_{P,Q}(z)&=\theta(P\ge 1)\,\theta(Q\ne 0)
\,\frac{(P+Q)!}{4}\sum_{k=0}^{P-1}
\Pmatrix{-Q\\ P-1-k}\frac{\sum_\pm\left(\pm z\right)^{k-P-Q}}{k!}
\nn\\
&\quad
-\left(\mbox{the same series with $P\rightleftarrows Q$}\right)
\qquad\qquad
\left(P+Q=0,1,2\right) ,
\label{eq:F_PQ(z)}
\end{align}
where $\theta(P\ge 1)$ and $\theta(Q\ne 0)$ are defined by
\begin{equation}
\theta(\textrm{condition})=
\begin{cases}
1 & \mbox{if the condition is satisfied}
\\
0 & \mbox{otherwise}
\end{cases}
.
\label{eq:theta(cond)}
\end{equation}
Then, $S_{m_1,m_2,m_3}$ is given as
\begin{align}
S_{m_1,m_2,m_3}
&=m_1f_{m_1+1,m_2+m_3}+m_2f_{m_2+1,m_3+m_1}+m_3f_{m_3+1,m_1+m_2}
\nn\\
&\quad
-\left(m_2+m_3\right)f_{m_2+m_3+1,m_1}
-\left(m_3+m_1\right)f_{m_3+m_1+1,m_2}
-\left(m_1+m_2\right)f_{m_1+m_2+1,m_3} ,
\label{eq:S_by_f}
\end{align}
where $f_{P,Q}$ is
\begin{equation}
f_{P,Q}=F_{P,Q}(2\pi i) .
\label{eq:f_PQ=F_PQ(2pii)}
\end{equation}
Note that $F_{P,Q}(z)$ and hence $f_{P,Q}$ are anti-symmetric with
respect to $(P,Q)$.

In summary, we have shown that $\calN[\alpha_k]$ is given by
a series of equations; \eqref{eq:calN_by_calN_n1n2n3}, 
\eqref{eq:calN_n1n2n3_by_S_m1m2m3}, \eqref{eq:S_by_f},
\eqref{eq:f_PQ=F_PQ(2pii)} and \eqref{eq:F_PQ(z)}.
It is a polynomial in $z^2=(2\pi i)^2$ and, as shown in Appendix
\ref{app:calN=N+}, the $z^0$ term is equal to $N$:
\begin{equation}
\calN[\alpha_k]=N+O(z^2) .
\label{eq:calN=N+O(z^2)}
\end{equation}
The terms of non-trivial power of $z^2$ are the ``anomalous'' part.
We present the analysis of the anomalous part as well as that of
$\calT[\alpha_k]$ in Sec.\ \ref{sec:Sol_with_calN=N_calT=0} after
obtaining a calculable expression of $\calT[\alpha_k]$ in the next
section.

\section{Expression of $\calT[\alpha_k]$}
\label{sec:calT}

First, let us express the EOM test $\calT$ \eqref{eq:calT}
in terms of $E_{abc}$.
From the $\Ke$-regularized version of \eqref{eq:UQBU^-1} and
\eqref{eq:(UQBU^-1)^2},
\begin{align}
\bigl(\Psie\bigr)_{14}
&=(E_{124})_\veps\left(c\Ke\right)_{12}\left(Bc\right)_{24} ,
\label{eq:Psie=}
\\
\bigl(\Psie^2\bigr)_{14}
&=\left(E_{124}-E_{134}\right)_\veps\left(c\Ke\right)_{12}
\left(c\Ke\right)_{23}\left(Bc\right)_{34} ,
\end{align}
we find that the EOM is violated (apparently) by $O(\veps)$:
\begin{align}
\bigl(\QB\Psie+\Psie^2\bigr)_{14}
&=(E_{124})_\veps\left[\left(cKc\Ke\right)_{12}\left(Bc\right)_{24}
-\left(c\Ke\right)_{12}\left(cKBc\right)_{24}\right]
\nn\\
&\quad
+\left(E_{124}-E_{134}\right)_\veps\left(c\Ke\right)_{12}
\left(c\Ke\right)_{23}\left(Bc\right)_{34}
\nn\\
&=\veps\times (E_{124})_\veps\left(c\Ke\right)_{12}c_{24} .
\label{eq:EOMviolation}
\end{align}
From this and \eqref{eq:Psie=}, $\calT$ is given by
\begin{equation}
\calT=\veps\int\!(T_{1234})_\veps
\left(c\Ke\right)_{12}\left(Bc\right)_{23}
\left(c\Ke\right)_{34}c_{41} ,
\label{eq:calT=}
\end{equation}
with $T_{1234}$ defined by
\begin{equation}
T_{1234}=E_{123}E_{341} .
\label{eq:T_1234=E_123E_341}
\end{equation}
As in the previous section, all the quantities in the rest of this
section should be regarded as $\Ke$-regularized ones, and we omit
the corresponding subscript $\veps$. For example, $T_{1234}$ means
$(T_{1234})_\veps$.

For evaluating $\calT$ which is multiplied by $\veps$, it is
sufficient to take the leading part of the expansion
\eqref{eq:Expand_E_abc} of $E_{123}$ in inverse powers of $G_a$.
In the present calculation, $G_a$ is simply
\begin{equation}
G(\Ke)=\frac{1}{\Ke} .
\label{eq:G(Ke)_for_calT}
\end{equation}
Using \eqref{eq:Expand_E_abc} and keeping only the leading part,
we see that $T_{1234}$ \eqref{eq:T_1234=E_123E_341} is given by
\begin{align}
&T_{1234}=\sum_{k=0}^N\alpha_k\left[
\left(\frac{G_1}{G_3}\right)^k-\left(\frac{G_1}{G_2}\right)^k
-\left(\frac{G_1}{G_2}\right)^N\left(\frac{G_2}{G_3}\right)^k
\right]
\nn\\
&\qquad\qquad\times
\sum_{\ell=0}^N\alpha_\ell\left[
\left(\frac{G_3}{G_1}\right)^\ell-\left(\frac{G_3}{G_4}\right)^\ell
-\left(\frac{G_3}{G_4}\right)^N\left(\frac{G_4}{G_1}\right)^\ell
\right] .
\end{align}
Substituting this into \eqref{eq:calT=}, we find that $\calT$ is given
in terms of $\calT_{n_1,n_2,n_3,n_4}$ defined by
\begin{align}
\calT_{n_1,n_2,n_3,n_4}&=\veps\int\!
G_1^{n_1}G_2^{n_2}G_3^{n_3}G_4^{n_4}
\left(c\Ke\right)_{12}\left(Bc\right)_{23}
\left(c\Ke\right)_{34}c_{41}
\nn\\
&=\veps\int\!Bc\frac{1}{\Ke^{n_3}}c\frac{1}{\Ke^{n_4-1}}
c\frac{1}{\Ke^{n_1}}c\frac{1}{\Ke^{n_2-1}} ,
\label{eq:calT_n1n2n3n4}
\end{align}
as
\begin{align}
\calT[\alpha_k]&=\sum_{k,\ell=0}^N\alpha_k\alpha_\ell\Bigl[
\calT_{k-\ell,0,\ell-k,0}-\calT_{k,0,\ell-k,-\ell}-\calT_{\ell-k,0,k,-\ell}
-\calT_{k-\ell,-k,\ell,0}+\calT_{k,-k,\ell,-\ell}
\nn\\
&\qquad\qquad
+\calT_{k-\ell,-k,N,\ell-N}-\calT_{\ell,-k,k-\ell,0}
+\calT_{N,k-N,\ell-k,-\ell}+\calT_{\ell,-k,k,-\ell}
\Bigr] ,
\label{eq:calT_by_calT_n1n2n3n4}
\end{align}
where we have made the replacement of the summation indices
$(k,\ell)\to(N-k,N-\ell)$ for the third, the seventh and the last
terms on the RHS to eliminate $N$ from the indices.
Therefore, the calculation of $\calT$ is reduced to that of
$\calT_{n_1,n_2,n_3,n_4}$. Note that the indices of
$\calT_{n_1,n_2,n_3,n_4}$ appearing in
\eqref{eq:calT_by_calT_n1n2n3n4} satisfy
\begin{equation}
n_1+n_2+n_3+n_4=0 .
\label{eq:n1+n2+n3+n4=0}
\end{equation}
In Appendix \ref{app:calT_n1n2n3n4_by_h}, we show that
$\calT_{n_1,n_2,n_3,n_4}$ is given as
\begin{align}
\calT_{n_1,n_2,n_3,n_4}
&=n_1\left(h_{n_3+n_4-1}-h_{n_3}-h_{n_4-1}\right)
+n_3\left(h_{n_4+n_1-1}-h_{n_4-1}- h_{n_1}\right)
\nn\\
&\quad
+\left(n_4-1\right)
\left(h_{n_3+n_4}+h_{n_4+n_1}-h_{n_4}-h_{n_3+n_4+n_1}\right) ,
\label{eq:calT_n1n2n3n4_by_h}
\end{align}
where $h_Q$ for an integer $Q$ is given by
\begin{equation}
h_Q=H_Q(2\pi i) ,
\label{eq:h_Q=H_Q(2pii)}
\end{equation}
with $H_Q(z)$ (which is not the Hermite polynomial) defined by
\begin{align}
H_Q(z)&=\sum_\pm\frac{1}{\pm z}\left[
\theta(Q\le -2)\sum_{k=0}^{-Q-2}\Pmatrix{-Q\\ k+2}
\frac{\left(\pm z\right)^k}{k!}
-\theta(Q\ge 1)\sum_{k=0}^{Q-1}\Pmatrix{Q+1\\ k+2}
\frac{\left(\pm z\right)^k}{k!}\right] .
\label{eq:H_Q(z)_by_series}
\end{align}
Note that $H_Q(z)$ is a polynomial in $z^2$. 

\section{Solutions with $\calN=N$ and $\calT=0$}
\label{sec:Sol_with_calN=N_calT=0}

Having obtained calculable expressions of $\calN[\alpha_k]$ and
$\calT[\alpha_k]$ in Secs.\ \ref{sec:calN} and \ref{sec:calT},
respectively, we in this section examine whether CSFT solutions
satisfying both of
\begin{equation}
\calN[\alpha_k]=N,
\qquad
\calT[\alpha_k]=0 ,
\label{eq:calN=N_calT=0}
\end{equation}
exist, namely, whether there exists $\{\alpha_k\}$ satisfying the two
conditions of \eqref{eq:calN=N_calT=0} for each $N$.
Of course, it is desirable to present a general argument applicable to
any $N$. However, we have not yet succeeded in keeping the complicated
expressions of $\calN[\alpha_k]$ and $\calT[\alpha_k]$ under full
control sufficient for general arguments.
Postponing complete analysis to future studies, we here present
arguments for various values of $N$.

As independent elements among $\alpha_k$ ($k=0,1,\cdots, N$) subject
to the constraints \eqref{eq:sum_kalpha_k=1} and
\eqref{eq:alpha_k=alpha_N-k}, we take the first $[N/2]$ ones,
$(\alpha_0,\alpha_1,\cdots,\alpha_{[N/2]-1})$.
Since our solution for $N=1$ is unique,
$(\alpha_0,\alpha_1)=(1/2,1/2)$, and agrees with that of
\cite{ES09} satisfying $\calN=1$ and $\calT=0$, let us start with the
$N=2$ case. In the following, $z^2$ implies $(2\pi i)^2$.

\subsection{$\alpha_k$ for $N=2,3,4,5$}

\noindent
\underline{$N=2$}\\
For $N=2$, $\calN$ and $\calT$ are given by
\begin{equation}
\calN=2+\alpha_0 z^2,
\qquad
\calT=-12\alpha_0 .
\end{equation}
Therefore, $\calN=2$ and $\calT=0$ are simultaneously realized by
taking $\alpha_0=0$:
\begin{equation}
\left(\alpha_0,\alpha_1,\alpha_2\right)
=(0,1,0) .
\label{eq:alpha_N=2}
\end{equation}
In this case, $\pF_{ab}$ \eqref{eq:pF_ab=} and $E_{abc}$
\eqref{eq:E_abc} for the $3$-brane solution are
\begin{equation}
\pF_{ab}=1-G_aG_b,
\qquad
E_{abc}=1-\frac{G_a+G_c}{G_b}+\frac{1}{G_b^2} .
\label{eq:pF_E_for_3-brane}
\end{equation}
Explicitly, the solution is given by
\begin{equation}
\Psi_\textrm{3-brane}=cK\!\left(1+\frac{1}{G^2}\right)\!Bc
-GcK\frac{1}{G}Bc-cK\frac{1}{G}Bc\,G ,
\label{eq:Psi_3-brane}
\end{equation}
with $G(K)$ of \eqref{eq:G_fixed}.

\noindent
\underline{$N=3$}\\
For $N=3$, we obtain
\begin{equation}
\calN=3+3\left(\alpha_0+\frac16\right)z^2,
\qquad
\calT=-36\left(\alpha_0+\frac16\right) .
\end{equation}
$\calN=3$ and $\calT=0$ are simultaneously realized by taking
$\alpha_0=-1/6$:
\begin{equation}
\left(\alpha_0,\alpha_1,\alpha_2,\alpha_3\right)
=\left(-\frac16,\frac23,\frac23,-\frac16\right) .
\end{equation}

\noindent
\underline{$N=4$}\\
For $N=4$, we obtain
\begin{align}
\calN&=4+\left(1+8\alpha_0+2\alpha_1\right)z^2+\frac12\alpha_0\,z^4 ,
\nn\\
\calT&=-12\left(1+8\alpha_0+2\alpha_1\right)
-\frac{10}{3}\alpha_0\,z^2 .
\end{align}
Demanding $\calN=4$ and $\calT=0$, $\alpha_k$ are uniquely determined
by two equations,
$1+8\alpha_0+2\alpha_1=0$ and $\alpha_0=0$, as
\begin{equation}
\left(\alpha_0,\alpha_1,\alpha_2,\alpha_3,\alpha_4\right)
=\left(0,-\frac12,2,-\frac12,0\right) .
\end{equation}

\noindent
\underline{$N=5$}\\
For $N=5$, we have
\begin{align}
\calN&=5+\frac52\left(1+6\alpha_0+2\alpha_1\right)z^2
+\frac12\left(6\alpha_0+\alpha_1\right)z^4 ,
\nn\\
\calT&=-30\left(1+6\alpha_0+2\alpha_1\right)
-\frac{10}{3}\left(6\alpha_0+\alpha_1\right)z^2 ,
\end{align}
and the conditions $\calN=5$ and $\calT=0$ uniquely determine
$\alpha_k$ as
\begin{equation}
\left(\alpha_0,\alpha_1,\alpha_2,\alpha_3,\alpha_4,\alpha_5\right)
=\left(\frac16,-1,\frac43,\frac43,-1,\frac16\right) .
\end{equation}

\subsection{$\alpha_k$ for $N\ge 6$}
As seen above, $\calN$ and $\calT$ for $N\le 5$ are polynomials in
$z^2$ and take the following form:
\begin{align}
\calN&=N+\sum_{n=1}^{[N/2]}f_n(\alpha_k)\,z^{2n} ,
\nn\\
\calT&=-\sum_{n=1}^{[N/2]}t_n f_n(\alpha_k)\,z^{2(n-1)} ,
\label{eq:calN=N+sum_calT=sum}
\end{align}
where $f_n(\alpha_k)$ are linear functions of $\alpha_k$
($k=0,1,\cdots,[N/2]-1$), and $t_n$ in $\calT$
are numerical coefficients.
In particular, $f_n(\alpha_k)$ are common between $\calN$ and $\calT$.
As we see below, the form \eqref{eq:calN=N+sum_calT=sum} is valid for
larger $N$ we will test. It must be possible to prove
\eqref{eq:calN=N+sum_calT=sum} for a generic $N$ by using
the expressions of $\calN$ and $\calT$ given in Secs.\ \ref{sec:calN}
and \ref{sec:calT}.

Then, a problem with the $N\ge 6$ cases is that, while the number of
conditions is only two of \eqref{eq:calN=N_calT=0},
the number of independent $\alpha_k$ is $[N/2]$, which is greater than
$2$ for $N\ge 6$. 
A general solution $\{\alpha_k\}$ to \eqref{eq:calN=N_calT=0} contains
powers of $\pi^2$ and is generically irrational.
In order to fix $\alpha_k$ uniquely, we here adopt a special (and
probably a ``natural'') solution to \eqref{eq:calN=N_calT=0} by
demanding $[N/2]$ conditions,
\begin{equation}
f_n(\alpha_k)=0\quad (n=1,\cdots,[N/2]) .
\label{eq:f_n=0}
\end{equation}
In fact, $\{\alpha_k\}$ given above for $N=2,3,4,5$ have been
determined by \eqref{eq:f_n=0}.
For larger $N$, the conditions \eqref{eq:f_n=0} provide us with
sufficient conditions to uniquely determine $\{\alpha_k\}$,
and the resultant $\alpha_k$ is a rational number, as far as we have
checked.
Here, we present $\calN$ and $\calT$ and the solution to
\eqref{eq:f_n=0} in the cases $N=6$, $7$ and $11$, as examples.

\noindent
\underline{$N=6$}\\
In this case, $\calN$ and $\calT$ are certainly of the form of
\eqref{eq:calN=N+sum_calT=sum}:
\begin{align}
\calN&=6+\left(4+27\alpha_0+12\alpha_1+3\alpha_2\right)z^2
+\frac12\left(21\alpha_0+6\alpha_1+\alpha_2\right)z^4
+\frac{1}{24}\alpha_0 z^6 ,
\nn\\
\calT&=-12\left(4+27\alpha_0+12\alpha_1+3\alpha_2\right)
-\frac{10}{3}\left(21\alpha_0+6\alpha_1+\alpha_2\right)z^2
-\frac{7}{30}\alpha_0 z^4 .
\end{align}
The solution to \eqref{eq:f_n=0} is given by:
\begin{equation}
\left(\alpha_0,\alpha_1,\alpha_2,\alpha_3,
\alpha_4,\alpha_5,\alpha_6\right)
=\left(0,\frac23,-4,\frac{23}{3},-4,\frac23,0\right) .
\end{equation}

\noindent
\underline{$N=7$}\\
In this case also, $\calN$ and $\calT$ are of the form of
\eqref{eq:calN=N+sum_calT=sum}:
\begin{align}
\calN&=7+7\left(1+6\alpha_0+3\alpha_1+\alpha_2\right)z^2
+\frac14\left(1+110\alpha_0+40\alpha_1+10\alpha_2\right)z^4
+\frac{1}{24}\left(8\alpha_0+\alpha_1\right)z^6 ,
\nn\\
\calT&=-84\left(1+6\alpha_0+3\alpha_1+\alpha_2\right)
-\frac53\left(1+110\alpha_0+40\alpha_1+10\alpha_2\right)z^2
-\frac{7}{30}\left(8\alpha_0+\alpha_1\right)z^4 .
\end{align}
The solution to \eqref{eq:f_n=0} is
\begin{equation}
\left(\alpha_0,\alpha_1,\alpha_2,\alpha_3,
\alpha_4,\alpha_5,\alpha_6,\alpha_7\right)
=\left(-\frac{3}{10},\frac{12}{5},-\frac{32}{5},\frac{24}{5},
\frac{24}{5},-\frac{32}{5},\frac{12}{5},-\frac{3}{10}\right) .
\end{equation}

\noindent
\underline{$N=11$}\\
In the case of $N=11$, $\calN$ and $\calT$ are of the form
\eqref{eq:calN=N+sum_calT=sum} with $f_n(\alpha_k)$ and
$t_n$ given by
\begin{align}
f_1&=11\left(\frac52+15\alpha_0+10\alpha_1+6\alpha_2
+3\alpha_3+\alpha_4\right),
\nn\\
f_2&=\frac34\left(9+510\alpha_0+290\alpha_1+150\alpha_2+66\alpha_3
+20\alpha_4\right),
\nn\\
f_3&=\frac{7}{24}\left(\frac{1}{14}+113\alpha_0
+47\alpha_1+17\alpha_2+5\alpha_3+\alpha_4\right),
\nn\\
f_4&=\frac{1}{720}\left(220\alpha_0+55\alpha_1+10\alpha_2
+\alpha_3\right),
\nn\\
f_5&=\frac{1}{40320}\left(12\alpha_0+\alpha_1\right),
\end{align}
and
\begin{equation}
\left(t_1,t_2,t_3,t_4,t_5\right)
=\left(12,\frac{20}{3},\frac{28}{5},\frac{36}{7}\right) .
\end{equation}
The solution to \eqref{eq:f_n=0} is
\begin{equation}
\left(\alpha_0,\alpha_1,\alpha_2,\alpha_3,\alpha_4\right)
=\left(-\frac{691}{210},\frac{1382}{35},-\frac{20528}{105},
\frac{10652}{21},-\frac{24384}{35}\right) .
\end{equation}

Summarizing this section, as far as we have checked for various
positive integer $N$,
$\calN$ and $\calT$ take the form of \eqref{eq:calN=N+sum_calT=sum} in
terms of common linear functions $f_n(\alpha_k)$, and the condition
\eqref{eq:f_n=0} uniquely determines $\{\alpha_k\}$.\footnote{
  We have checked this for $N$ up to $35$ by using Mathematica.
  }
Of course, there are many questions to be answered and subjects to be
studied, which we shall discuss in the next section.

\section{Summary and discussions}
\label{sec:summary}

In this paper, we have presented an analytic expression of the
multi-brane solutions of CSFT for arbitrary (positive integer) brane
numbers.
We started with the most generic unitary and real string field $U$
\eqref{eq:U_by_Gm_pF} with $\Gm_a$ and $\pF_{ab}$ satisfying 
\eqref{eq:pF_aa=1-Gm_a^2} and \eqref{eq:pF_ab=pF_ba},
and considered as a candidate solution the pure-gauge
string field $U\QB U^{-1}$. As $\Gm_a$ and $\pF_{ab}$ for multi-brane
solutions, we adopted the ansatz of \eqref{eq:Gm_a=} and
\eqref{eq:pF_ab=} using $G(K)$ with a simple pole at $K=0$.
For the $(N+1)$-brane solution, we in this paper demanded the
following two:
First, the energy density of the solution calculated from the action
should be that of the $(N+1)$-brane. Concretely, $\calN$
\eqref{eq:calN} should be equal to the integer $N$.
Second, the EOM test against the solution itself given by $\calT$
\eqref{eq:calT} should vanish.
In the previous constructions of multi-brane solutions based on the
singularity at $K=0$, these two conditions were hard to be realized in
the cases of $N\ge 2$.
In the present construction, our solution is specified by real
parameters $\alpha_k$ subject to \eqref{eq:sum_kalpha_k=1} and
\eqref{eq:alpha_k=alpha_N-k}, and the problem is whether there exists
$\{\alpha_k\}$ which realizes $\calN=N$ and $\calT=0$.
We calculated $\calN[\alpha_k]$ and $\calT[\alpha_k]$ in the
$\Ke$-regularization to find that there indeed exists $\{\alpha_k\}$
satisfying the two conditions for any $N=2,3,4,5,\cdots$
as far as we have tested.
For $N\ge 6$, the two conditions, $\calN=N$ and $\calT=0$, cannot
uniquely fix $\alpha_k$, and we proposed to demand stronger
conditions \eqref{eq:f_n=0} on $\alpha_k$, which give sufficient
number of equations to uniquely determine $\alpha_k$ as rational
numbers.

Here, we add a remark for preventing a possible
misunderstanding of the reader about our construction of
solutions.
One might think that our construction is almost trivial and
meaningless since we are imposing only the two conditions 
\eqref{eq:calN=N_calT=0} on the solutions, and this is always possible
if the candidate solution has enough number of parameters ($\alpha_k$
in our case).
However, we should recall that our candidate solution is ``almost a
solution'' since it is of the pure-gauge form $\Psi=U\QB U^{-1}$,
which automatically satisfies the EOM if there is no subtlety at
$K=0$.
The non-integer nature of $\calN$ and, possibly, the failure of the
EOM test against itself, for a generic $\{\alpha_k\}$
would be manifestations of the non-regularity of $U$ at $K=0$
as we explained in the Introduction.
The two conditions we impose should be regarded as conditions
necessary for making the pure-gauge configuration a more regular one.

We have certainly succeeded in constructing $(N+1)$-brane solutions
satisfying the two conditions for $N=2, 3, 4, 5,\cdots$.
However, our analysis in this paper is still at an ``experimental''
level.
Namely, we have confirmed the existence of the ``natural'' choice
of $\{\alpha_k\}$ determined by \eqref{eq:f_n=0} only for sample
values of $N$.
Although there is no doubt that such $\{\alpha_k\}$ giving a desired
multi-brane solution exists for any integer $N$, we should present a
general proof for our expectation. For this, we have to show that the
expressions of $\calN$ and $\calT$ given in
\eqref{eq:calN=N+sum_calT=sum} in terms of common functions
$f_n(\alpha_k)$ are valid for any $N$.
It is of course desirable that the solution $\alpha_k$ to
\eqref{eq:f_n=0} is explicitly given for a generic $N$.

Even if these technical problems are resolved, there still remain
important questions on our construction of multi-brane solutions:
\begin{itemize}
\item
What is the meaning of the stronger conditions \eqref{eq:f_n=0} on
$\alpha_k$?
Possibly, these conditions could be derived by considering other
natural requirements on the solution. For example, the requirement
that the energy density of the solution evaluated from the
gravitational coupling \cite{AHandNI,GRSZ,Ellwood,KKT}
be equal to that of the $(N+1)$-brane. Besides, since the
number of conditions \eqref{eq:f_n=0} depends on $N$ (and is equal to
$[N/2]$), requirements related to the fluctuation modes on the
solution might be the origins of the conditions.

\item
Is there any profound mathematical meaning in $\pF_{ab}$ given by 
\eqref{eq:pF_ab=} in terms of $\alpha_k$ satisfying the condition
\eqref{eq:f_n=0}?
Recalling that $\calN$ \eqref{eq:calN} for the present pure-gauge type
solution $\Psi=U\QB U^{-1}$ has an analogy to the winding number
$\calW[g]$ \eqref{eq:calW} of the mapping $g(x)$ from a three-manifold
$M$ to a Lie group,
it would be interesting if the present construction realizing
arbitrary integer $\calN$ gives some hint for uncovering the meaning
of $\calN$ as ``winding number'' as we explained in the Introduction.

\item
In this paper, as the EOM tests, we considered only that against the
solution itself given by $\calT$ \eqref{eq:calT}.
Let us define the EOM test against a generic string field $\calO$
with $\Ngh=1$ by
\begin{equation}
\calT[\calO]=\int\!\calO*\left(\QB\Psie+\Psie^2\right) .
\label{eq:calT[calO]}
\end{equation}
It has been known that the $N=1$ (2-brane) solution does not pass the
EOM test against the Fock vacuum;
$\calT[(e^{-\frac{\pi}{4}K}c\,e^{-\frac{\pi}{4}K})_\veps]
=O(1/\veps)\ne 0$
\cite{MS2}.
This property also persists in our $N\ge 2$ solutions irrespective
of the choice of $\alpha_k$ as we have already mentioned in the
Introduction.
Instead, our solutions pass the EOM test against the unitary
transformed Fock vacuum;
$\calT[(Ue^{-\frac{\pi}{4}K}c\,e^{-\frac{\pi}{4}K}U^{-1})_\veps]=0$.
On the other hand, the tachyon vacuum solution ($N=-1$) passes all the
EOM tests.
For full understanding of the problem of the EOM test, it would be
necessary to solve the problem of the fluctuation modes around the
solution (see \cite{HataFluct}).
\end{itemize}

Among the above three questions/problems, the last one is the most
serious one from the viewpoint of constructing complete solutions.
However, we expect that, even if the third problem remains
unresolved, our finding in this paper gives a useful hint in
considering the topological aspects of CSFT as we stated in the
Introduction and in the above second question.
We finish this paper by giving some comments concerning our solution:
\begin{itemize}
\item In the particular case of $N=2$, our $U$ with $\alpha_k$ of
\eqref{eq:alpha_N=2} has the following manifestly unitary expression:
\begin{equation}
U=\exp\left(\frac12\bigl\{\CR{B}{c},g(K)\bigr\}\right) ,
\end{equation}
where $g(K)$ is defined by
\begin{equation}
e^{g(K)}=G(K)=\frac{1+K}{K} .
\end{equation}
In relation to this, the following $U$ is also unitary for any
self-conjugate $f(K)$:
\begin{equation}
U=\exp\bigl(f(K)\CR{B}{c}f(K)\bigr) .
\end{equation}
This $U$ is rewritten into the standard form \eqref{eq:U_by_Gm_pF} and
the corresponding $\Gm_a$ and $\pF_{ab}$ are
\begin{equation}
\Gm_a=e^{f(K_a)^2},
\qquad
\pF_{ab}=\frac{2\left(\ln\Gm_a\ln\Gm_b\right)^{1/2}}{
  \ln\Gm_a+\ln\Gm_b}\left(1-\Gm_a\,\Gm_b\right) .
\end{equation}
Note that this $\pF_{ab}$ is equal to $\pF_{ab}=1-\Gm_a\Gm_b$ in
\eqref{eq:pF_E_for_3-brane} for $N=2$ (recall that $\Gm=G$ when $N=2$)
multiplied by the front term consisting of $\ln\Gm$.
However, we find that, due to the presence of the $\ln\Gm$ term in
$\pF_{ab}$, both $\calN$ and $\calT$ are {\em divergent} in the limit
$\veps\to +0$. In fact, if we take
$\Gm=G(K)=(1+K)/K$, $\calN$ diverges as
\begin{equation}
\calN=O\!\left(\frac{1}{\veps^2\ln^2(1/\veps)}\right) .
\end{equation}
In this respect also, our $\pF_{ab}$ given by \eqref{eq:pF_ab=} is
a good choice.

\item
The product $U^{(3)}=U^{(1)}U^{(2)}$ of two unitary $U^{(1)}$ and
$U^{(2)}$ is of course unitary and is written in the form
\eqref{eq:U_by_Gm_pF} with $\Gm_a$ and $\pF_{ab}$ satisfying 
\eqref{eq:pF_aa=1-Gm_a^2} and \eqref{eq:pF_ab=pF_ba}.
In fact, $(\Gm_a,\pF_{ab})$ of $U^{(3)}$ is given in terms of those of
$U^{(1)}$ and $U^{(2)}$ by
\begin{equation}
\Gm_a^{(3)}=\Gm_a^{(1)}\Gm_a^{(2)},
\qquad
\pF^{(3)}_{ab}=\pF^{(1)}_{ab}+\Gm^{(1)}_a\pF^{(2)}_{ab}\Gm^{(1)}_b .
\end{equation}
This relation implies that, even if $(\Gm_a^{(1,2)},\pF_{ab}^{(1,2)})$
are of the form of \eqref{eq:Gm_a=} and \eqref{eq:pF_ab=}, 
$(\Gm_a^{(3)},\pF_{ab}^{(3)})$ is no longer so and cannot realize
integer $\calN=N^{(3)}=N^{(1)}+N^{(2)}$ and $\calT=0$ in general.
We have already seen this phenomenon of the violation of the
additivity of $\calN$ in the case of $N^{(1)}=N^{(2)}=1$ in
\cite{HKwn}.

\item
In this paper, we considered explicitly only $(N+1)$-brane solutions
with positive integer $N$. However, ``ghost brane'' solutions with
$N\le -2$ can also be constructed in the same manner.

\end{itemize}

\appendix

\section{Calculations for Sec.\  \ref{sec:MostGenUnitaryU} }
\label{app:MostGenUnitaryU}

In this Appendix, we present the derivations of some of the equations
in Sec.\ \ref{sec:MostGenUnitaryU}, in particular, 
the conditions \eqref{eq:pF_aa=1-Gm_a^2} and \eqref{eq:pF_ab=pF_ba}
for the unitarity of $U$ \eqref{eq:U_by_Gm_pF}.
Though the calculations are straightforward, they may be helpful as
examples of the convenient notation of this paper.

First, the conjugate of $U$ \eqref{eq:U_by_Gm_pF} is
\begin{equation}
\bigl(U^\dg\bigr)_{12}=\frac{1}{\Gm_1}\id_{12}
-\frac{\pF_{21}}{\Gm_1}(cB)_{12}
=\frac{1-\pF_{11}}{\Gm_1}\id_{12}+\frac{\pF_{21}}{\Gm_1}(Bc)_{12} ,
\label{eq:U^-1_calc}
\end{equation}
and $UU^\dg$ is given by
\begin{equation}
(U U^\dg)_{13}=U_{12}(U^\dg)_{23}
=\frac{1-\pF_{11}}{(\Gm_1)^2}\id_{13}
+\left[\frac{\left(1-\pF_{11}\right)\pF_{31}}{(\Gm_1)^2}
-\frac{\pF_{13}\left(1-\pF_{33}\right)}{(\Gm_3)^2}
\right](Bc)_{13} .
\label{eq:UU^dagger=}
\end{equation}
In deriving \eqref{eq:UU^dagger=}, we have used
\begin{equation}
f(K_2)\,(Bc)_{12}(Bc)_{23}=f(K_2)\,\id_{12}(Bc)_{23}
=f(K_1)\,(Bc)_{13} ,
\label{eq:Bc_Bc=}
\end{equation}
valid for any $f(K)$.
From \eqref{eq:UU^dagger=}, we find that $U$ is unitary if
the two conditions \eqref{eq:pF_aa=1-Gm_a^2} and
\eqref{eq:pF_ab=pF_ba} are satisfied.
Eq.\ \eqref{eq:U^-1=U^dagger=} for $U^{-1}$ follows immediately from
\eqref{eq:U^-1_calc} and the two conditions.

Next, let us evaluate $U\QB U^{-1}$ for $U$
\eqref{eq:U_by_Gm_pF} and $U^{-1}$ \eqref{eq:U^-1=U^dagger=}:
\begin{equation}
\left(U\QB U^{-1}\right)_{13}
=U_{12}\left(\QB U^{-1}\right)_{23}
=\left[\frac{1}{\Gm_2}\id_{12}-\frac{\pF_{12}}{\Gm_2}(Bc)_{12}\right]
\frac{\pF_{23}}{\Gm_2}\,(cKBc)_{23}
=E_{123}\left(cK\right)_{12}\left(Bc\right)_{23} ,
\label{eq:UQBU^-1_deriv}
\end{equation}
where $E_{abc}$ is given by \eqref{eq:E_abc}.
In the calculation of \eqref{eq:UQBU^-1_deriv}, we have used
$\QB(Bc)=cKBc$, the identity
\begin{equation}
(Bc)_{12}\left(cKBc\right)_{23}
=\id_{12}\left(cKBc\right)_{23}
-\left(cK\right)_{12}\left(Bc\right)_{23} ,
\end{equation}
or more generally,
\begin{equation}
(Bc)_{12}\left(cK\right)_{23}\left(Bc\right)_{34}
=\Bigl(\id_{12}\left(cK\right)_{23}
-\left(cK\right)_{12}\id_{23}\Bigr)\left(Bc\right)_{34} ,
\label{eq:Bc_cK_Bc=}
\end{equation}
and the condition \eqref{eq:pF_aa=1-Gm_a^2}.

\section{The formula \eqref{eq:delta_calN=}}
\label{app:Deriv_deltacalN}

For an arbitrary infinitesimal deformation $\delta U^{-1}$, we have
\begin{equation}
\delta\!\left(U\QB U^{-1}\right)=\QB\!\left(U\delta U^{-1}\right)
+\CR{U\QB U^{-1}}{U\delta U^{-1}} .
\end{equation}
Using this, we obtain
\begin{align}
\frac{1}{\pi^2}\,\delta\calN
&=\int\!\left(U\QB U^{-1}\right)_\veps^2
\,\delta\!\left(U\QB U^{-1}\right)_\veps
=\int\!\left(U\QB U^{-1}\right)_\veps^2
\left\{\QB\!\left(U\delta U^{-1}\right)
+\CR{U\QB U^{-1}}{U\delta U^{-1}}\right\}_\veps
\nn\\
&=\int\left\{\left(U\QB U^{-1}\right)^2
\QB\!\left(U\delta U^{-1}\right)\right\}_\veps
=\int\!\left\{\QB\!\left[\left(U\QB U^{-1}\right)^2
\left(U\delta U^{-1}\right)\right]\right\}_\veps ,
\end{align}
where we have used $\QB(U\QB U^{-1})^2=0$ in obtaining the last
expression.
Then, noticing that $\bigl(\QB f(K)\bigr)_\veps=0=\QB f(\Ke)$,
$(\QB c)_\veps=c\Ke c=cKc=\QB c$
and $(\QB B)_\veps=\Ke=\QB B+\veps$,
we see that the following relation holds for any $\calO(K,B,c)$:
\begin{equation}
\bigl(\QB\calO(K,B,c)\bigr)_\veps
=\QB\calO(\Ke,B,c)+\veps\times T_B\calO(\Ke,B,c) ,
\end{equation}
where $T_B$ is the operation \eqref{eq:T_B}.
Since $\int\!\QB\calO(\Ke,B,c)$ vanishes without ambiguity, we obtain
\eqref{eq:delta_calN=}.

\section{Derivation of \eqref{eq:S_by_f}}
\label{app:S_by_F}

In this Appendix, we derive eq.\ \eqref{eq:S_by_f} for
$S_{m_1,m_2,m_3}$ \eqref{eq:S_m1m2m3} by using the $(s,z)$-integration
formula for the $Bcccc$-correlators \cite{MS1,MS2}.
This formula is given by
\begin{equation}
\int\!BcF_1(K)cF_2(K)cF_3(K)cF_4(K)
=\int_0^\infty\!ds\,\frac{s^2}{(2\pi)^3\,i}
\int_{-i\infty}^{i\infty}\!\frac{dz}{2\pi i}\,e^{sz}\,
\calG(z),
\label{eq:MSformula}
\end{equation}
where $\calG(z)$ is defined in our convention by
\begin{align}
\calG(z)
&=\Bigl[(\Ds F_1)F_2F_3'+F_1'F_2(\Ds F_3)
+\bigl(F_1(\Ds F_2)F_3\bigr)'
-\Ds(F_1F_2')\,F_3
\nn\\
&\qquad
-\Ds(F_1F_2)\,F_3'
-F_1'\Ds(F_2F_3)-F_1\Ds(F_2'F_3)
+\Ds(F_1F_2'F_3)\Bigr]F_4 ,
\end{align}
with $F_i=F_i(z)$, $F_i'=(d/dz)F_i(z)$ and
\begin{equation}
(\Ds F_i)(z)\equiv F_i\!\left(z-\frac{2\pi i}{s}\right)
-F_i\!\left(z+\frac{2\pi i}{s}\right) .
\end{equation}

In the application of \eqref{eq:MSformula} to the $ccc$-correlator in
$S_{m_1,m_2,m_3}$ \eqref{eq:S_m1m2m3}, $F_i(z)$ are
(note that $cBc=c$)
\begin{equation}
F_a(z)=\frac{1}{\left(z+u+\veps\right)^{m_a}}
\quad\left(a=1,2,3\right),
\qquad
F_4(z)=1 .
\label{eq:F_a_calN_org}
\end{equation}
In this case, the contour of $z$-integration \eqref{eq:MSformula} can
be closed by adding the infinite semi-circle in the left-half plane
$\Re z<0$ due to the presence of $e^{sz}$.
In addition, we find that the infinitesimal positive constant $\veps$
in $S_{m_1,m_2,m_3}$ \eqref{eq:S_m1m2m3} is totally absorbed into the
following replacements (rescaling) of the three integration variables
$(u,s,z)$:
\begin{equation}
\left(u,s,z\right)\to\left(\veps u,\frac{s}{\veps},\veps z\right) .
\end{equation}
Then, we obtain
\begin{equation}
\frac{1}{\pi^2}\,S_{m_1,m_2,m_3}
=\int_0^\infty\!\!\frac{du}{(2\pi i)^3}\,u^{1+\sum_a m_a}
\int_0^\infty\!ds\,s^2\,\sum_{\parbox{11.1mm}{
    \scriptsize poles in\\ $\Re z<0$}}
\Res e^{sz}\,\calG(z) ,
\label{eq:S_m1m2m3_by_dudsRes}
\end{equation}
where $F_i(z)$ for the present $\calG(z)$ is given, instead of
\eqref{eq:F_a_calN_org}, by
\begin{equation}
F_a(z)=\frac{1}{\left(z+u+1\right)^{m_a}}
\quad\left(a=1,2,3\right),
\qquad
F_4(z)=1 .
\end{equation}

Explicitly, $\calG(z)$ in \eqref{eq:S_m1m2m3_by_dudsRes} is given by
\begin{align}
\calG(z)&=\sum_{\pm}(\pm)\Biggl\{
-\frac{m_3}{\wh{z}^{\;m_2+m_3+1}\left(\wh{z}\mp\da\right)^{m_1}}
-\frac{m_1}{\wh{z}^{\;m_1+m_2+1}\left(\wh{z}\mp\da\right)^{m_3}}
-\frac{m_3+m_1}{\wh{z}^{\;m_3+m_1+1}\left(\wh{z}\mp\da\right)^{m_2}}
\nn\\
&\qquad
-\frac{m_2}{\wh{z}^{\;m_3+m_1}\left(\wh{z}\mp\da\right)^{m_2+1}}
+\frac{m_2}{\wh{z}^{\;m_3}\left(\wh{z}\mp\da\right)^{m_1+m_2+1}}
+\frac{m_3}{\wh{z}^{\;m_3+1}\left(\wh{z}\mp\da\right)^{m_1+m_2}}
\nn\\
&\qquad
+\frac{m_1}{\wh{z}^{\;m_1+1}\left(\wh{z}\mp\da\right)^{m_2+m_3}}
+\frac{m_2}{\wh{z}^{\;m_1}\left(\wh{z}\mp\da\right)^{m_2+m_3+1}}
-\frac{m_2}{\left(\wh{z}\mp\da\right)^{m_1+m_2+m_3+1}}
\Biggr\} ,
\label{eq:calG_for_S}
\end{align}
where $\wh{z}$ is defined by $\wh{z}\equiv z+u+1$.
Note that the contribution of each term in \eqref{eq:calG_for_S} to
$S_{m_1,m_2,m_3}$ \eqref{eq:S_m1m2m3_by_dudsRes} is given by the
following $f_{P,Q}$:
\begin{equation}
\frac{1}{\pi^2}\,f_{P,Q}
=\sum_\pm(\pm)
\int_0^\infty\!\!\frac{du}{(2\pi i)^3}\,u^{P+Q}
\int_0^\infty\!ds\,s^2\sum_{\parbox{11.1mm}{
    \scriptsize poles in\\ $\Re z<0$}}
\Res\frac{e^{sz}}{\left(z+u+1\right)^P\left(z+u+1\mp\da\right)^Q}
\label{eq:f_PQ=}
\end{equation}
where a pair of integers $(P,Q)$ satisfy
\begin{equation}
P+Q=\sum_{a=1}^3 m_a+1=0, 1, 2 .
\end{equation}
Calculating the sum of residues in \eqref{eq:f_PQ=} at
$z=-u-1$ and $-u-1\pm(2\pi i/s)$ by using the formulas,
\begin{align}
\Res_{z=0}\,\frac{e^{sz}}{z^m\left(z+a\right)^n}
&=\theta(m\ge 1)
\sum_{k=0}^{m-1}\frac{1}{k!}\Pmatrix{-n\\ m-1-k}
s^k a^{k-n-m+1}
\qquad\left(n\ne 0\right) ,
\label{eq:Res_formula1}
\\[3mm]
\Res_{z=0}\frac{e^{sz}}{z^{m}}
&=\theta(m\ge 1)\frac{s^{m-1}}{(m-1)!} ,
\label{eq:Res_formula2}
\end{align}
and carrying out the $u$- and $s$-integrations,
we find that $f_{P,Q}$ is given by \eqref{eq:f_PQ=F_PQ(2pii)}
by using $F_{P,Q}(z)$ \eqref{eq:F_PQ(z)}, and that $S_{m_1,m_2,m_3}$
is given by \eqref{eq:S_by_f} in terms of $f_{P,Q}$ (by using the
anti-symmetry of $f_{P,Q}$ for a number of terms). 
In particular, the last term of \eqref{eq:calG_for_S} does not
contribute to $S_{m_1,m_2,m_3}$ since we have $f_{P,0}=f_{0,Q}=0$
(the residues cancel after the summation $\sum_\pm(\pm$)).

The series $F_{P,Q}(z)$ \eqref{eq:F_PQ(z)} has the following
expression in terms of the confluent hypergeometric functions:
\begin{align}
F_{P,Q}(z)&=\theta(P\ge 1)\,\theta(Q\ne 0)
\times\left(-\frac14\right)\sum_\pm
\begin{cases}
\ds
Q\,{}_1F_1(Q+1,2;\pm z) & (P+Q=0)
\\[2mm]
\ds
\frac{1}{(\pm z)}\,{}_1F_1(Q,1;\pm z) & (P+Q=1)
\\[4mm]
\ds
\frac{2}{(\pm z)}\,{}_1F_1(Q,2;\pm z) & (P+Q=2)
\end{cases}
\nn\\
&\quad
-\left(P\rightleftarrows Q\right) .
\label{eq:F_PQ(z)_by_1F1}
\end{align}

\section{Proof of  \eqref{eq:calN=N+O(z^2)}}
\label{app:calN=N+}

As we saw in Sec.\ \ref{sec:calN[alpha_k]}, $\calN$ for our solution
is given as a polynomial in $z^2=(2\pi i)^2$.
In this Appendix, we show \eqref{eq:calN=N+O(z^2)}, namely, that the
$z^0$ term of $\calN$ is equal to $N$.

Let us start with the expression \eqref{eq:F_PQ(z)} of $F_{P,Q}(z)$.
Here, we repeatedly use the fact that the integers $P$ and $Q$ for
$F_{P,Q}(z)$ are restricted to those of the three cases:
\begin{equation}
P+Q=0,1,2 .
\label{eq:P+Q=0,1,2}
\end{equation}
First, note that the negative power terms of $z$ in \eqref{eq:F_PQ(z)}
are actually non-existent. Next,
since the $z^0$ part of \eqref{eq:F_PQ(z)} comes from the $k=P+Q$
term, and this term is in the range of the $k$-summation only when
$P+Q\le P-1$, namely, $Q\le -1$, we obtain
\begin{align}
F_{P,Q}(z=0)&=\theta(P\ge 1)\,\theta(Q\le -1)\,\frac14
\Pmatrix{-Q\\ -Q-1}\sum_\pm 1-(P\rightleftarrows Q)
\nn\\
&=-\theta(P\ge 1)\,\theta(Q\le -1)\,\frac{Q}{2}
+\theta(Q\ge 1)\,\theta(P\le -1)\,\frac{P}{2} .
\end{align}
Then, by taking into account \eqref{eq:P+Q=0,1,2}, we find that
$F_{P,Q}(0)$ is rewritten as
\begin{equation}
F_{P,Q}(0)=-\theta(P\ge 1)\,\frac{Q}{2}
+\theta(Q\ge 1)\,\,\frac{P}{2}
=\theta(Q\ge 1)\,\frac{P+Q}{2}-\frac{Q}{2}
-\frac12\,\theta(P=1)\,\theta(Q=1) ,
\label{eq:F_PQ(0)}
\end{equation}
where we need \eqref{eq:P+Q=0,1,2} also at the second equality.
Plugging this into $S_{m_1,m_2,m_3}$ \eqref{eq:S_by_f} given by
$f_{P,Q}$, we find that the contribution of the $-Q/2$ term of
\eqref{eq:F_PQ(0)} 
cancels, while the $-(1/2)\theta(P=1)\,\theta(Q=1)$ term does not
contribute since $f_{P,Q}$ is multiplied by $P-1$ in
\eqref{eq:S_by_f}.
Therefore, we get
\begin{align}
S_{m_1,m_2,m_3}\bigr|_{z=0}
&=\frac{\sum_a m_a+1}{2}\Bigl[
\theta(m_2+m_3\ge 1)\,m_1+\theta(m_3+m_1\ge 1)\,m_2
+\theta(m_1+m_2\ge 1)\,m_3
\nn\\
&\quad
-\theta(m_1\ge 1)\left(m_2+m_3\right)
-\theta(m_2\ge 1)\left(m_3+m_1\right)
-\theta(m_3\ge 1)\left(m_1+m_2\right)
\Bigr] .
\end{align}
Plugging this into $\calN_{n_1,n_2,n_3}$
\eqref{eq:calN_n1n2n3_by_S_m1m2m3} given by $S_{m_1,m_2,m_3}$, and
using that $\sum_{a=1}^3 n_a=1$, we obtain
\begin{align}
\calN_{n_1,n_2,n_3}\bigr|_{z=0}
&=-\theta(n_1\ge 1)-\theta(n_2\ge 1)-\theta(n_3\ge 1)+1
\nn\\
&=-\theta(n_1\ge 1)\,\theta(n_2\ge 1)
-\theta(n_2\ge 1)\,\theta(n_3\ge 1)
-\theta(n_3\ge 1)\,\theta(n_1\ge 1) ,
\label{eq:calN_n1n2n3|z=0}
\end{align}
where, in the derivation of the first expression, we have used, for
example,
\begin{equation}
\theta(n_2+n_3\ge 1)-\theta(n_2+n_3\ge 2)
=\theta(n_2+n_3=1)=\theta(n_1=0) .
\end{equation}
Finally, substituting \eqref{eq:calN_n1n2n3|z=0} into $\calN$
\eqref{eq:calN_by_calN_n1n2n3}
and using $\calN_{n_1,0,n_3}\bigr|_{z=0}=0$, we get
\begin{equation}
\calN\bigr|_{z=0}=2\sum_{k,\ell=0}^N\alpha_k\alpha_\ell
\Bigl[\ell\,\theta(k\ge\ell)+\left(N-\ell\right)\theta(k\ge\ell+1)
\Bigr] .
\end{equation}
Making the replacement of the summation indices
$(k,\ell)\to(N-k,N-\ell)$ for the second term,
we obtain
\begin{equation}
\calN\bigr|_{z=0}=2\sum_{k,\ell=0}^N\alpha_k\alpha_\ell
\,\ell\bigl[\theta(k\ge\ell)+\theta(k\le\ell-1)\bigr]
=2\sum_{k=0}^N\alpha_k\sum_{\ell=0}^N\ell\alpha_\ell
=N ,
\end{equation}
where we have used \eqref{eq:sum_kalpha_k=1} and
\eqref{eq:sum_kkalpha_k=N/2}.

\section{Derivation of
  \eqref{eq:calT_n1n2n3n4_by_h} -- \eqref{eq:H_Q(z)_by_series}} 
\label{app:calT_n1n2n3n4_by_h}

The $Bcccc$-correlator $\calT_{n_1,n_2,n_3,n_4}$
\eqref{eq:calT_n1n2n3n4} is evaluated by using the formula
\eqref{eq:MSformula}. In this case, the four functions $F_i(z)$ are
\begin{equation}
F_1(z)=\frac{1}{\left(z+\veps\right)^{n_3}},
\quad
F_2(z)=\frac{1}{\left(z+\veps\right)^{n_4-1}},
\quad
F_3(z)=\frac{1}{\left(z+\veps\right)^{n_1}},
\quad
F_4(z)=\frac{1}{\left(z+\veps\right)^{n_2-1}} .
\end{equation}
Taking into account \eqref{eq:n1+n2+n3+n4=0}, we see that the
positive infinitesimal constant $\veps$ in $\calT_{n_1,n_2,n_3,n_4}$
can be absorbed into the rescaling of two integration
variables $(s,z)$,
\begin{equation}
\left(s,z\right)\to\left(\frac{s}{\veps},\veps z\right) ,
\end{equation}
to obtain
\begin{equation}
\calT_{n_1,n_2,n_3,n_4}
=-\frac{1}{(2\pi i)^3}\int_0^\infty\!ds\,s^2
\sum_{\parbox{11.1mm}{
    \scriptsize poles in\\ $\Re z<0$}}
\Res e^{sz}\,\calG(z) ,
\label{eq:calT=intdsRes}
\end{equation}
where $\calG(z)$ is given by
\begin{align}
\calG(z)
&=\sum_\pm(\pm)\biggl\{
-\frac{n_1}{\wh{z}^{\;n_4+n_1+n_2-1}\left(\wh{z}\mp\da\right)^{n_3}}
-\frac{n_3}{\wh{z}^{\;n_2+n_3+n_4-1}\left(\wh{z}\mp\da\right)^{n_1}}
\nn\\
&\quad
-\frac{n_1+n_3}{\wh{z}^{\;n_1+n_2+n_3}\left(\wh{z}\mp\da\right)^{n_4-1}}
-\frac{n_4-1}{\wh{z}^{\;n_1+n_2+n_3-1}\left(\wh{z}\mp\da\right)^{n_4}}
+\frac{n_4-1}{\wh{z}^{\;n_1+n_2-1}\left(\wh{z}\mp\da\right)^{n_3+n_4}}
\nn\\
&\quad
+\frac{n_1}{\wh{z}^{\;n_1+n_2}\left(\wh{z}\mp\da\right)^{n_3+n_4-1}}
+\frac{n_3}{\wh{z}^{\;n_2+n_3}\left(\wh{z}\mp\da\right)^{n_4+n_1-1}}
+\frac{n_4-1}{\wh{z}^{\;n_2+n_3-1}\left(\wh{z}\mp\da\right)^{n_4+n_1}}
\nn\\
&\quad
-\frac{n_4-1}{\wh{z}^{\;n_2-1}\left(\wh{z}\mp\da\right)^{n_3+n_4+n_1}}
\biggr\} ,
\label{eq:calG(z)_calT}
\end{align}
with $\wh{z}\equiv z+1$.
Note that each term in \eqref{eq:calG(z)_calT} is of the form
$1/\bigl(\wh{z}^{\,P}\left(\wh{z}\mp 2\pi i/s\right)^Q\bigr)$
with $P+Q=\sum_{a=1}^4n_a-1=-1$.
Therefore, defining $h_Q$ by
\begin{equation}
h_Q=-\frac{1}{(2\pi i)^3}\sum_\pm(\pm)
\int_0^\infty\!ds\,s^2\sum_{\parbox{11.1mm}{
    \scriptsize poles in\\ $\Re z<0$}}
\Res \frac{e^{sz}}{\left(z+1\right)^{-Q-1}\left(z+1\mp\da\right)^Q} ,
\label{eq:h_Q=}
\end{equation}
we see that $\calT_{n_1,n_2,n_3,n_4}$ is given in terms of $h_Q$ by
\eqref{eq:calT_n1n2n3n4_by_h}.
For $h_Q$ \eqref{eq:h_Q=}, calculating the sum of residues at $z=-1$
and $-1\pm(2\pi i/s)$ by using the formula \eqref{eq:Res_formula1}
and carrying out the $s$-integration, we obtain
\begin{align}
h_Q&=-\frac{1}{2\pi i}\sum_\pm(\pm)\biggl[
\theta\left(-Q-1\ge 1\right)\sum_{k=0}^{-Q-2}\Pmatrix{-Q\\ k+2}
\frac{\left(\mp 2\pi i\right)^k}{k!}
\nn\\
&\hspace{4.5cm}
+\theta\left(Q\ge 1\right)\sum_{k=0}^{Q-1}\Pmatrix{Q+1\\ k+2}
\frac{\left(\pm 2\pi i\right)^k}{k!}\biggr] .
\end{align}
This leads to the expression of $h_Q$ given by 
\eqref{eq:h_Q=H_Q(2pii)} and \eqref{eq:H_Q(z)_by_series}.
The series $H_Q(z)$ \eqref{eq:H_Q(z)_by_series} is expressed by the
confluent hypergeometric functions as
\begin{equation}
H_Q(z)=-\frac{Q(Q+1)}{2}\frac{1}{z}\sum_\pm(\pm)
\Bigl[\theta(Q\le -2)\,{}_1F_1(2+Q,3;\pm z)
-\theta(Q\ge 1)\,{}_1F_1(1-Q,3;\pm z)\Bigr] .
\label{eq:H_Q(z)_CHGF}
\end{equation}


\end{document}